\documentclass[aps,twocolumn,showpacs,preprintnumbers,amsmath,amssymb,nofootinbib,showkeys]{revtex4-1}

\usepackage{hyperref}
\usepackage{float}
\usepackage{epsfig}
\usepackage{graphicx}
\usepackage{mathptmx}      
\usepackage{latexsym}
\usepackage{natbib}
\usepackage{longtable}
\usepackage{dcolumn}
\usepackage[utf8]{inputenc}
\usepackage{amsmath}
\usepackage{url}
\usepackage{xcolor}

\def\XXint#1#2#3{{\setbox0=\hbox{$#1{#2#3}{\int}$ }
\vcenter{\hbox{$#2#3$ }}\kern-.6\wd0}}

\def\1{\'{\i}}
\def\S{{\rm Short}}

\DeclareMathVersion{nxbold}
\SetSymbolFont{operators}{nxbold}{OT1}{cmr} {b}{n}
\SetSymbolFont{letters}  {nxbold}{OML}{cmm} {b}{it}
\SetSymbolFont{symbols}  {nxbold}{OMS}{cmsy}{b}{n}

\DeclareMathAlphabet{\mathcal}{OMS}{cmsy}{m}{n}
\begin{document}

\title{Low energy peripheral scaling in Nucleon-Nucleon
  scattering and uncertainty quantification}


\author{I. Ruiz Simo}
\email{ruizsig@ugr.es}
\author{J.E. Amaro}
\email{amaro@ugr.es}
\author{E. Ruiz Arriola}
\email{earriola@ugr.es}
\affiliation{Departamento de F\'isica At\'omica, Molecular y Nuclear and Instituto Carlos I de
  Fisica Te\'orica y Computacional,  Universidad de Granada, E-18071  Granada, Spain}
\author{R. Navarro P\'erez}
\email{navarrop@ohio.edu}
\affiliation{Institute of Nuclear and Particle Physics and Department of
Physics and Astronomy, \\ Ohio University, Athens, OH 45701, USA}

\date{\today}


\begin{abstract}
We analyze the peripheral structure of the nucleon-nucleon interaction
for LAB energies below 350 MeV. To this end we transform the
scattering matrix into the impact parameter representation by
analyzing the scaled phase shifts $(L+1/2) \delta_{JLS} (p)$ and the
scaled mixing parameters $(L+1/2)\epsilon_{JLS}(p)$ in terms of the
impact parameter $b=(L+1/2)/p$. According to the eikonal
approximation, at large angular momentum $L$ these functions should
become an universal function of $b$, {\it independent} on $L$.  This
allows to discuss in a rather transparent way the role of statistical
and systematic uncertainties in the different long range components of
the two-body potential.  Implications for peripheral waves obtained in
chiral perturbation theory interactions to fifth order (N5LO) or from
the large body of NN data considered in the SAID partial wave analysis
are also drawn from comparing them with other phenomenological
high-quality interactions, constructed to fit scattering data as well.
We find that both N5LO and SAID peripheral waves disagree more than $5
\sigma$ with the Granada-2013 statistical analysis, more than $ 2
\sigma$ with the 6 statistically equivalent potentials fitting the
Granada-2013 database and about $1 \sigma $ with the historical set of
13 high-quality potentials developed since the 1993 Nijmegen analysis.
\end{abstract}

\keywords{NN interaction, nucleon-nucleon phase-shifts, 
peripheral scaling, long-range correlations 
}
\pacs{03.65.Nk,11.10.Gh,13.75.Cs,21.30.Fe,21.45.+v}

\maketitle

\section{Introduction}

The analysis of NN scattering has been a field of intensive research
since it provides a good starting point to constrain NN interactions
in Nuclear Physics.  The mid-range distance region, that proves to be
crucial for nuclear binding studies, can so far most accurately be
determined by direct fits to NN scattering data. However,
unfortunately, this region is not tightly constrained by low energy
scattering data, typically below $T_{\rm LAB}=350$ MeV.  Moreover, the
inclusion of well-known long distance effects, such as charge
dependent one pion exchange (CD-OPE) interaction, Coulomb, vacuum
polarization, relativistic and magnetic moments effects prove crucial
for extracting the needed mid-range component from standard
$\chi^2$-fits to the abundant np and pp scattering data at those
energies. These fits provide statistically significant confidence that
the difference between theory and experiment is a fluctuation whose
nature (usually a gaussian) can be determined. This requirement to
validate the partial wave analysis (PWA) has been emphasized since the
early days~(see
e.g. \cite{Breit:1962zz,arndt1966chi,signell1969nuclear} for reviews
and references therein) and was scrupulously followed by the Nijmegen
group~\cite{Stoks:1993tb} and the subsequent NijmI, NijmII,
Reid93~\cite{Stoks:1994wp}, AV18~\cite{Wiringa:1994wb}, CD
Bonn~\cite{Machleidt:2000ge} and Spectator~\cite{Gross:2008ps}
potentials. The statistical high-quality of these 7 nuclear potentials
---potentials with a high-statistical confidence and a corresponding
$\chi^2 /\nu \sim 1 $--- was possible due to the implementation of the
small but crucial long-distance effects mentioned above. These effects
were missing in many previous analyses and lead to low statistical
significance~\cite{Stoks:1993zz,Stoks:1994pi}. The most recent
analysis~\cite{Perez:2013mwa,Perez:2013jpa} accomplished a
$3\sigma$-selfconsistent selected pp+np database involving 6713 data
and normalizations in the LAB energy range between $1$ eV for np and
$338$ KeV for pp and a maximum of $350$ MeV for both np and
pp\footnote{The Granada-2013 database can be downloaded from the
  website \url{http://www.ugr.es/~amaro/nndatabase/}.}. This has lead
to the new 6 Granada potentials denoted as
DS-OPE~\cite{Perez:2013mwa,Perez:2013jpa},
DS-$\chi$TPE~\cite{Perez:2013oba,Perez:2013cza},
SOG-OPE~\cite{Perez:2014yla}, SOG-$\chi$TPE, DS-$\Delta$BO and
SOG-$\Delta$BO~\cite{Perez:2014waa}.

In this paper we will focus in analyzing thoroughly the long distance
behavior of the NN interaction characterized by peripheral scattering.
We will also consider some informative tests which set important and
tight constraints on phase-shifts with large angular momentum on the
light of these 13 high-quality analyses, sharing {\it exactly} the
same long distance CD-OPE and electromagnetic interactions.

Let us first review the long distance structure of the NN potential in
a way that our problem can be easily formulated. The functional form
of the NN interaction reflects the exchange of the lightest mesons by
Yukawa-like interactions.  The time-honored OPE potential dominates
the longest distance for $r > 3$ fm, and it is given by
\begin{eqnarray}
 V_{{\rm OPE},pp}(r) &=& f^2_{p} V_{m_{\pi^0},\rm OPE}(r) \, ,\label{ope1} \\
 V_{{\rm OPE},nn}(r) &=& f^2_{n} V_{m_{\pi^0},\rm OPE}(r) \, ,\label{ope2}  \\
 V_{{\rm OPE},np}(r) &=& -f_{n}f_{p}V_{m_{\pi^0},\rm OPE}(r) \nonumber \\  &-& (-)^{T} 2f^2_cV_{m_{\pi^\pm},\rm OPE}(r) \, , 
 \label{eq:BreakIsospinOPE}
\end{eqnarray}
 where $T$ is the isospin of the np pair. Here $V_{m,\rm OPE}$ is given by
\begin{equation}
 V_{m, \rm OPE}(r) = \left(\frac{m}{m_{\pi^\pm}}\right)^2\frac{1}{3}m\left[Y_m(r){\mathbf \sigma}_1\cdot\mathbf{\sigma}_2 + T_m(r)S_{12} \right] \, , 
\end{equation}
being $Y_m$ and $T_m$ the usual Yukawa functions, 
\begin{eqnarray}
 Y_m(r) &=& \frac{e^{-m r}}{m r} \, , \\  
 T_m(r) &=& \frac{e^{-m r}}{m r } \left[ 1 + \frac3{mr} + \frac{3}{(mr)^2}\right] \, , 
\end{eqnarray}
where $\sigma_1$ and $\sigma_2$ are the single nucleon Pauli matrices
and $S_{12}= 3 \sigma_1 \cdot {\bf \hat r} \sigma_2 \cdot {\bf \hat r}
-\sigma_1 \cdot \sigma_2 $ is the tensor operator.

While the pion masses $m$ can be directly determined independently of
the NN interaction, the couplings in the CD-OPE potential need
consideration of $\pi N$ or $NN$ scattering processes, which in turn
require a PWA. Based on the Granada-2013
database~\cite{Perez:2013jpa}, the most accurate determination of the
couplings $f_N$ has been reported by a recent
analysis~\cite{Arriola:2016hfi,Perez:2016aol} with
$\chi^2/\nu=1.025$. One of the reasons why the couplings can be
determined so accurately is the fact that tiny changes in the tail
provide important modifications in the small angle and low energy
observables, which have been accurately measured. This long distance
component of the interaction is very compelling but it is formulated
in coordinate space and hence it is not directly accessible to
experimental determination. Another possibility is given by an {\it ab
  initio} computation of the static energy between two point sources
made of three quarks with nucleon quantum numbers on the QCD
lattice~\cite{Aoki:2011ep,Aoki:2013tba}, but so far this approach
still provides much larger uncertainties for the Yukawa couplings
compared with phenomenological approaches based on a PWA.

\begin{figure*}
\centering
\epsfig{file=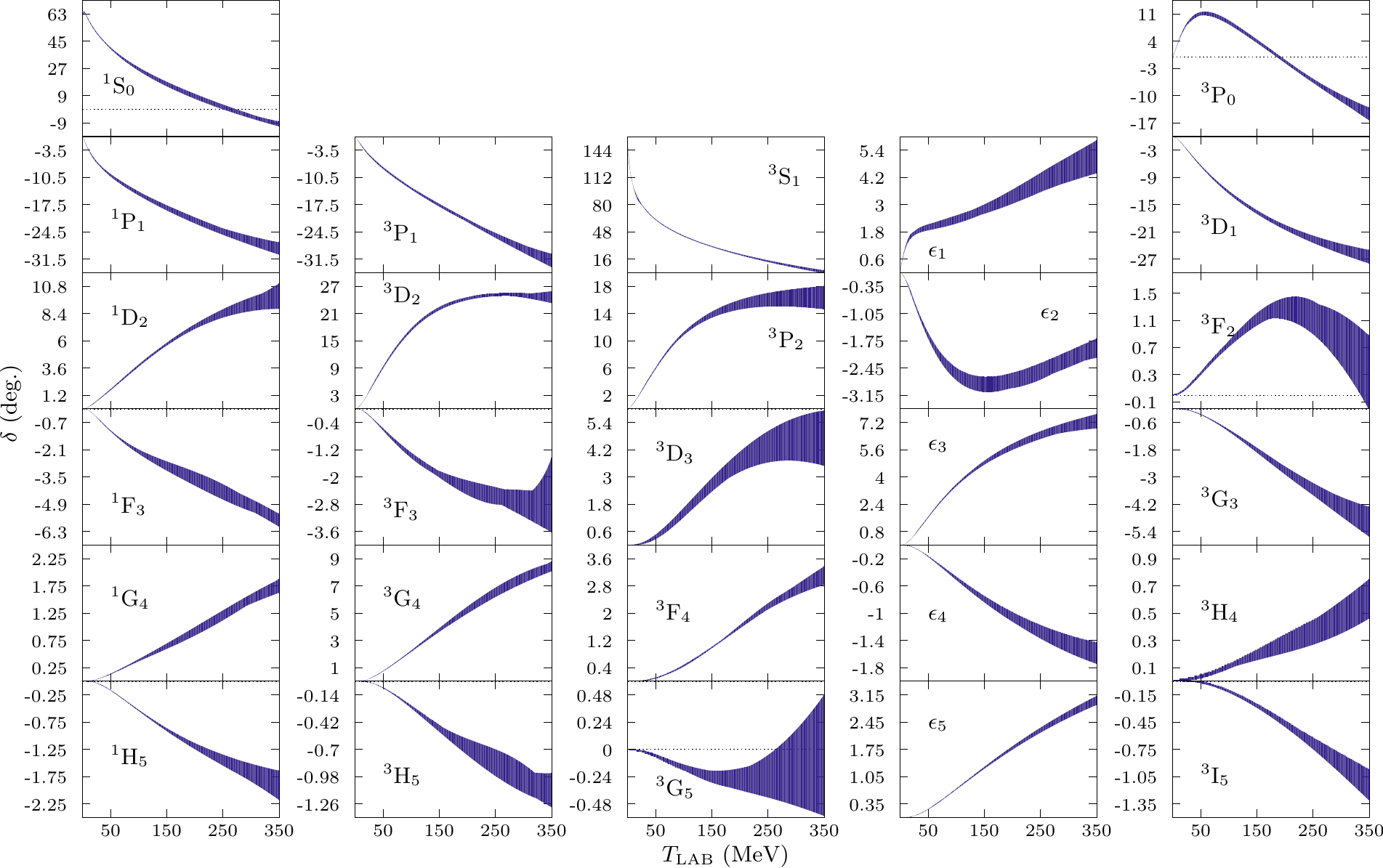,width=18cm}
\caption{Neutron-proton phase-shifts (and mixing angle) bands between
  $ {\rm min}\, \delta_i$ and ${\rm max} \, \delta_i$ including 13
  high statistical quality potentials with the same CD-OPE tail in all
  partial waves with $J \le 5$ as a function of the LAB energy. We
  consider the Nijmegen PWA~\cite{Stoks:1993tb}, Nijm I, NijmII
  Reid93~\cite{Stoks:1994wp} the AV18~\cite{Wiringa:1994wb}, CD
  Bonn~\cite{Machleidt:2000ge}, Spectator~\cite{Gross:2008ps}, and the
  recent 6 Granada potentials denoted as
  DS-OPE~\cite{Perez:2013mwa,Perez:2013jpa},
  DS-$\chi$TPE~\cite{Perez:2013oba,Perez:2013cza},
  SOG-OPE~\cite{Perez:2014yla}, SOG-$\chi$TPE, DS-$\Delta$BO and
  SOG-$\Delta$BO~\cite{Perez:2014waa}.}
\label{fig:ps-sys}       
\end{figure*}

One way of highlighting the long distance information from NN
scattering is to look {\it a posteriori} at peripheral partial waves
with high angular momentum. The peripheral features of strong
interactions among elementary particles have been exploited for more
than half a century~\cite{okun1959peripheral} and they were
immediately applied to NN interactions in terms of OPE in the Born
approximation~\cite{signell1959nucleon,Cziffra:1959zza,breit1960note}. Tests
for NN peripherality and the meson exchange picture have been made at
the partial waves level for high angular
momenta~\cite{Binstock:1972gx}, and within the chiral perturbation
theory proposed by Weinberg~\cite{Weinberg:1990rz} at different
orders~\cite{Kaiser:1997mw,Kaiser:1998wa,Epelbaum:2003gr,Krebs:2007rh,Entem:2014msa}. One
drawback is that in perturbation theory the meson exchange picture
provides singular interactions at short distances, where the order of
the divergence generally increases with the order of the coupling
constant. However, for a given order there is a partial wave with
sufficiently high angular momentum where results are finite, so that
angular momentum acts as a perturbative regulator. This regularization
procedure is very appealing but it requires passing from the physical
and measurable momentum transfer $\vec q$ to the angular momentum
variable $L$, thus making necessary a PWA including both peripheral
and not peripheral partial waves. At this point it is worth reminding
that phase-shifts extracted from a PWA {\it are not} by themselves
observables at a given energy, unless a complete set of scattering and
polarization observables at that energy is available. Instead, one has
mostly incomplete measurements that demand some interpolation
procedure.

One should notice that all these peripheral waves studies are somewhat
qualitative and they are based on visual comparison to scattering
phases extracted from PWA. To our knowledge, the degree of agreement
or disagreement has never been quantified. As an illustration, in
Fig.~\ref{fig:ps-sys} we display np phase-shifts $\delta_{JLS}^{(i)}$
and mixing angles $\epsilon_{JLS}^{(i)}$ for all partial waves with $J
\le 5$ for 13 realistic interactions and in the conventional
fashion. We do so for 13 high-quality determinations, but instead of
plotting 13 different lines we present a band containing all of them,
namely for any energy we take the interval ${\rm min}_i
\delta^{(i)}_{JLS} \le \delta^{(i)}_{JLS} \le {\rm max}_i
\delta^{(i)}_{JLS} $ and similarly for $\epsilon^{(i)}_{JLS}$.  All
these interactions share the {\it same} CD-OPE tail as the dominant
feature above 3 fm, and the discrepancies reflect the differences in
the interactions below 3 fm, yet fitting the data in a statistically
significant manner by the time they were developed. Besides, from the
spread observed in all of them and the fact that higher partial waves
produce smaller phase-shifts, it is not obvious which $LSJ$ channels
can be compared for the {\it same} range of energies and to {\it what
  extent} are these differences significant.

The purpose of the present paper is threefold. First, we present a
{\it peripheral plot}, a universality pattern for the peripheral waves
in terms of the impact parameter, suggested by the eikonal
approximation.  This peripheral plot turns out to be very informative
regarding the relevant scales in NN interaction, and how many partial
waves are directly intertwined. This is in contrast to the usual plots
of phase-shifts in each partial wave as a function of the energy,
where this information is, in fact, hidden.  Second, we make a
clean-cut {\it quantitative} discussion on the role played by
different mid-range interactions on the higher $L$ partial
waves. Third, we visualize both statistical as well as systematic
errors (see also \cite{Perez:2014waa}), and we also single out which
phase-shifts behave as outliers when compared to well-stablished
high-quality potentials.

The paper is organized as follows.  In Section~\ref{sec:impact} we
discuss the meaning of "elementariness" in NN interactions as a help
to introduce the corresponding impact parameter.  In
Section~\ref{sec:peripheral} we introduce the peripheral plot in the
light of the eikonal approximation, which requires consideration of
coupled channels. We illustrate the scaling features of the OPE
interaction when treated in perturbation theory. In
Section~\ref{sec:peripheral-hq} we show that scaling indeed works for
realistic high-quality interactions when the uncertainties are taken
into account. In Section~\ref{sec:peripheral-test} we use the
peripheral plot as well as the corresponding statistical and
systematic uncertainties as a quantitative test for two important
determinations of peripheral phase-shifts.  Finally, in
Section~\ref{sec:concl} we summarize our results and present our main
conclusions. Technical details are further elaborated in
Appendices~\ref{sec:PW-conv}, \ref{sec:pert-ope} and \ref{sec:WKB}.

\section{Impact parameter and effective elementariness}
\label{sec:impact}

Meson exchange forces such as CD-OPE,
Eqs.~(\ref{ope1}-\ref{eq:BreakIsospinOPE}), play a similar role as van
der Waals interactions in molecular physics: they correspond to the
interaction between elementary point-like particles. Nucleons are
composite particles with size $\sim a$. This will modify
Eqs.~(\ref{ope1}-\ref{eq:BreakIsospinOPE}) for $r \lesssim r_c \sim 2
a$. For instance, the electromagnetic interaction between protons
requires consideration of charge distribution by means of a form
factor, but it reduces to the Coulomb potential $e^2/r$ for $r \gtrsim
1.8$ fm~\cite{Arriola:2016hfi}. Microscopically, regulated OPE and TPE
follow also a similar pattern, i.e, beyond an elementariness radius,
the interactions correspond to point-like particles.

\begin{table}
\centering 
\begin{tabular}{|c|c|c|}
\hline
$L $ & $T_{\rm LAB}^{\rm max}$ (MeV) & $T_{\rm LAB}^{\rm max}$ (MeV) \\
     & ($b_{\rm eff} >  r_c= 1.8 $fm) &  ($b_{\rm eff} >  r_c= 3 $fm) \\
\hline 
 0 & 6.4 & 2.3 \\
 1 & 57.5 & 20.7 \\
 2 & 159.9 & 57.6 \\
 3 & 313.5 & 112.9 \\
 4 & 518.3 & 186.6 \\
 5 & 774.2 & 278.7 \\
 6 & 1081.4 & 389.3 \\ 
\hline 
\end{tabular}
\caption{Maximum LAB energies for partial waves not intruding below
  the distances $b_{\rm eff} = b /\sqrt{2} > r_c= 1.8 {\rm fm}$ and
  $b_{\rm eff} > r_c= 3 {\rm fm}$}
\label{tab:impact}
\end{table}

In previous works we have advocated a coarse grained
approach~\cite{NavarroPerez:2011fm,NavarroPerez:2012qf,Perez:2013mwa}
for the unknown interaction below the elementariness radius $r_c= 3$
fm that gives the best quality fit to NN data below a laboratory (LAB)
energy of 350 MeV~\cite{Perez:2013jpa}. This maximum energy
corresponds to maximal CM momentum $p_{\rm max} \sim \sqrt{M m_\pi}$.
The coarse grain is based on the idea that, for a fixed angular
momentum $L$, the NN interaction is efficiently sampled at radial
points located at a relative distance $\Delta r \sim 1/p_{\rm max}$.

Within a semi-classical approach the relation between the CM momentum
$p$, the impact parameter $b$ and the angular momentum $L$ is given
by\footnote{Note that we are making $L(L+1) \to (L+1/2)^2$ motivated
  by the well-known Langer modification to ensure the correct $\sim
  r^{L+1}$ short distance behavior of the WKB wave function (see
  e.g.~\cite{pascualquantum}). }
\begin{eqnarray}
b p = L + \frac12 \, .  
\label{eq:impact}
\end{eqnarray}
Strictly speaking, the impact parameter is not an observable except at
high energies, but we can use it as a convenient variable. Note that
for a quantized $L$, and for a fixed energy, we have $\Delta b = 1/p$,
in agreement with the radial coarse grained interval $\Delta r$ .
Better semi-classical approximations provide suitable corrections to
the leading behavior, Eq.~(\ref{eq:impact}).  In any case,
Eq.~(\ref{eq:impact}) provides a sensible way to define when a given
partial wave can be regarded as peripheral, and this has to do with
the spanned range of impact parameters, where we know for certain that
the long range force is produced by pion exchange.

Based on the Granada-2013 np+pp database~\cite{Perez:2013jpa}, we
carried out comprehensive and statistically consistent fits allowing
to clearly identify the regions where there is only OPE (CD-OPE)
\cite{Perez:2013jpa,Arriola:2016hfi,Perez:2016aol} and one+two
pion-exchange (CD-OPE+TPE)
\cite{Perez:2013za,Perez:2013oba,Perez:2014bua}, namely
\begin{eqnarray}
V_{NN}(r) &=& V_{1\pi} (r) \, , \qquad \qquad \quad r > r_c= 3.0  \; {\rm fm} \\
V_{NN}(r) &=& 
V_{1\pi} (r) + V_{2 \pi} (r) \, , \qquad  r > r'_c= 1.8 \; {\rm fm}
\end{eqnarray}
For distances below $r'_c=1.8$ fm, finite nucleon size effects ($3\pi$
exchange, heavier mesons or quark exchange) become important and
difficult to disentangle. We effectively incorporate these short
distance components as a sum of delta-shells separated by $\Delta
r=0.6$ fm, with strengths fitted to the np and pp scattering
database. Attempts to extend TPE below that distance either produce
too large $\chi^2$ values or unnaturally large low energy constants,
and hence they had to be ruled
out~\cite{Perez:2014bua,RuizArriola:2016sbf}. Within the EFT framework
this means that counter-terms have a range smaller than $r'_c$.

\begin{figure}[t]
\centering
\includegraphics[width=8cm,height=7cm]{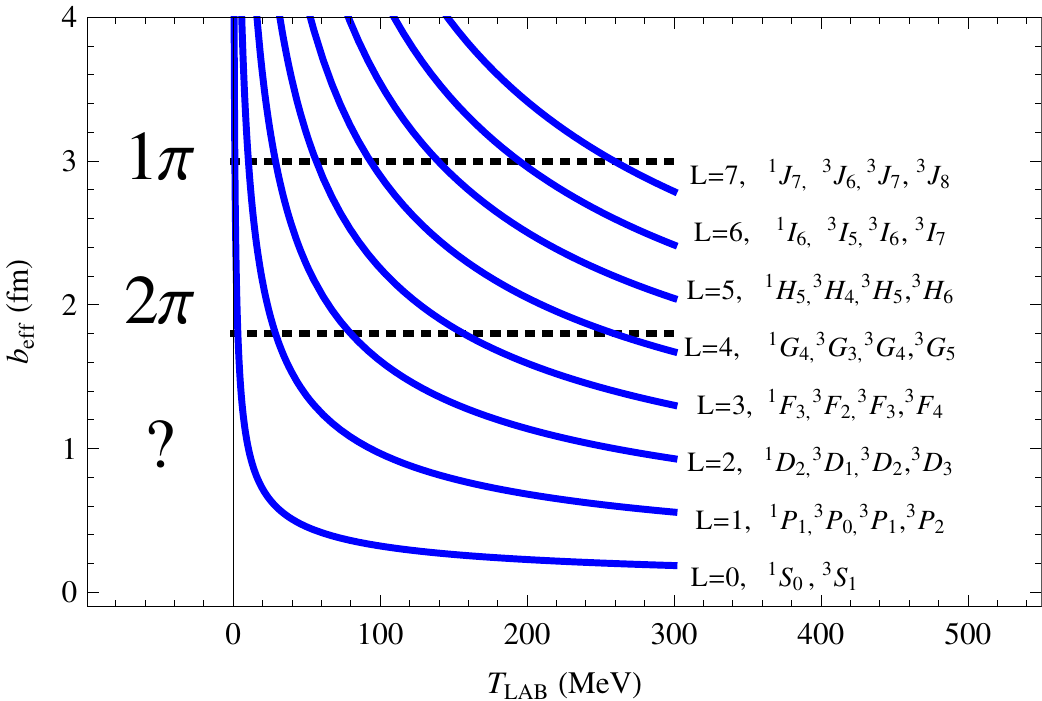}
\caption{Effective impact parameter $b_{\rm eff}= b/\sqrt{2}$ as a
  function of the LAB energy for different partial waves according to
  their angular momentum $L$ and the formula $ b p = (L+1/2)$ with
  $T_{\rm LAB}= 2 p^2 /M_N$. This probes the region of a short
  distance potential $V_\S(r)$ which vanishes {\it above} a certain
  distance (see main text). The horizontal dotted lines represent the
  distances above which the NN potential is described by pion
  exchanges in the Granada
  analyses~\cite{Perez:2013mwa,Perez:2013jpa,Perez:2013oba,Perez:2013cza,Perez:2014yla}. Here
  $V_{NN}(r) = V_{1\pi} (r) $ for $r > r_c= 3$ fm and $V_{NN}(r) =
  V_{1\pi} (r) + V_{2 \pi} (r) $ for $r > r'_c= 1.8$ fm.}
\label{fig:impact-TLAB}       
\end{figure}
The above discussion implies that the short distance component (the
delta-shells) of the potential has a sharp end, i.e. $V_\S (r) = 0$
for $r > r'_c$ \cite{Perez:2013za,Perez:2013oba,Perez:2014bua} or for
$r > r_c$ \cite{Perez:2013jpa,Arriola:2016hfi,Perez:2016aol},
depending on the analyses. If we denote the corresponding phase-shifts
by $\delta_{L,\S}(k)$, then the partial wave expansion of the
scattering amplitude corresponding to this short distance component
(we assume a central potential to simplify the discussion) can be
written as
\begin{eqnarray}
f_\S (\theta) = \sum_{L=0}^\infty ( 2 L+1) \frac{e^{2 i \delta_{L,\S}(k)}-1}{2 i k} P_L(\cos \theta) \, . \label{eq:scattering_amplitude}
\end{eqnarray}
The analogous expansion for the total cross section is then 
\begin{eqnarray}
\sigma_\S = \frac{4 \pi}{k^2}\sum_{L=0}^\infty (2 L+1) \sin^2 \delta_{L,\S} (k) \, .
\label{eq:xsect_expansion}
\end{eqnarray}
Based on Eq. (\ref{eq:impact}) with the impact parameter $b$ replaced
by the sharp end of the short distance component of the potential
($r'_c$ or $r_c$), one would expect that there is an effectively
maximum angular momentum for each energy, $L_{\rm max} \sim k r'_c$,
beyond which short distance phase-shifts become negligible, thus
truncating the infinite series of Eqs.
(\ref{eq:scattering_amplitude}) and (\ref{eq:xsect_expansion}).  While
this happens for the total cross section, it is not exactly the same
for the scattering amplitude $f_\S(\theta) $.  In fact, the scattering
amplitude at backward angles requires $L_{\rm max} \sim 2 k r'_c$ due
to a diffractive effect caused by the sharp boundary. These effects
are illustrated in Appendix \ref{sec:PW-conv} for a simple spherical
well potential.  Likewise, within a semi-classical context one should
have vanishing scattering for $b > r'_c$. Actually, this is not so,
and phase-shifts vanish only when $b \gtrsim \sqrt{2} r'_c$,
essentially due to a diffractive effect which vanishes at very short
wavelengths (see also Appendix \ref{sec:PW-conv} for the spherical
well potential case).  Therefore, it is useful to define an effective
impact parameter, $b_{\rm eff } = b /\sqrt{2}$, which complies better
with the actual situation.

We illustrate in Fig.~\ref{fig:impact-TLAB} the effective impact
parameter as a function of the LAB energy for fixed $L$ values
(Eq.~(\ref{eq:impact})). Some pertinent numerical values are presented
in Table~\ref{tab:impact}.

The cut radii for OPE and OPE+TPE are shown with horizontal lines.  As
we clearly see, below pion production threshold, $T_{\rm LAB} \le 290$
MeV, all waves with $L=0,1,2,3$ intrude into the short distance region
below the elementariness radius $r^\prime_c$. This implies that short
distance phenomenological components (or equivalently
EFT-counterterms, see below) for these partial waves will generally be
needed to describe scattering data in addition to the OPE and TPE
potentials.  Likewise, for this energy we may regard $G,H,I...$ waves
as truly peripheral.  If the LAB energy is reduced to $125$ MeV, then
the $F$ waves also behave as peripheral and they will need no short
distance phenomenological components.  This feature has been verified
explicitly by comparing the quality of the fits in different
scenarios~\cite{Perez:2014bua,RuizArriola:2016sbf}.

\begin{figure*}[hp]
\centering
\includegraphics[width=8cm]{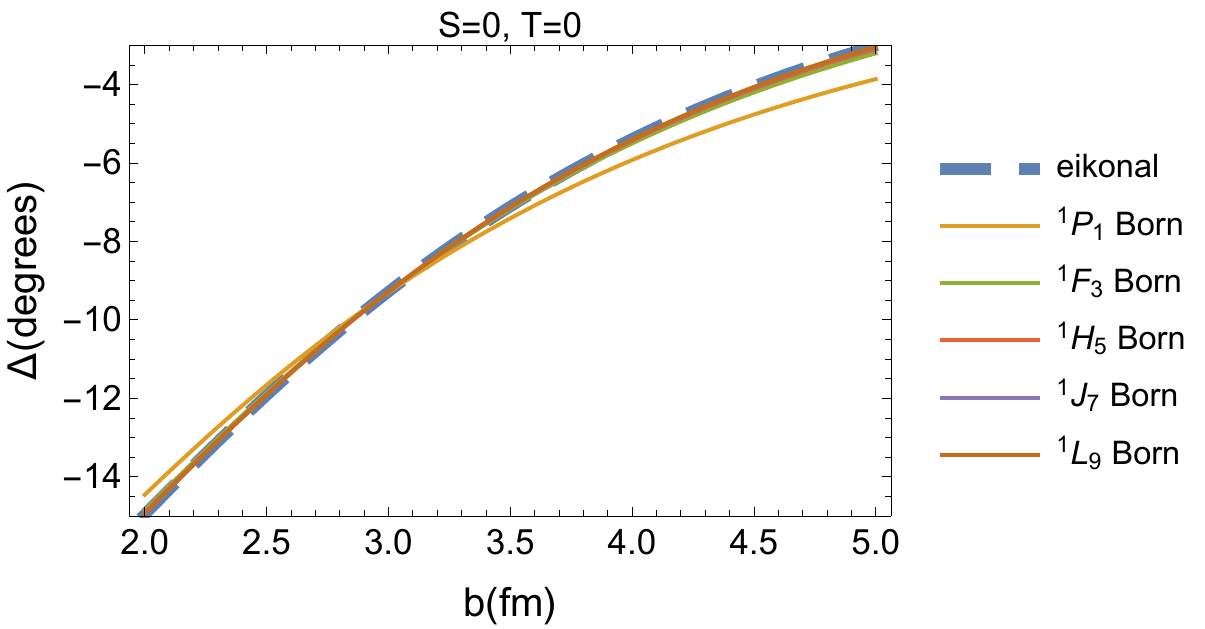}
\includegraphics[width=8cm]{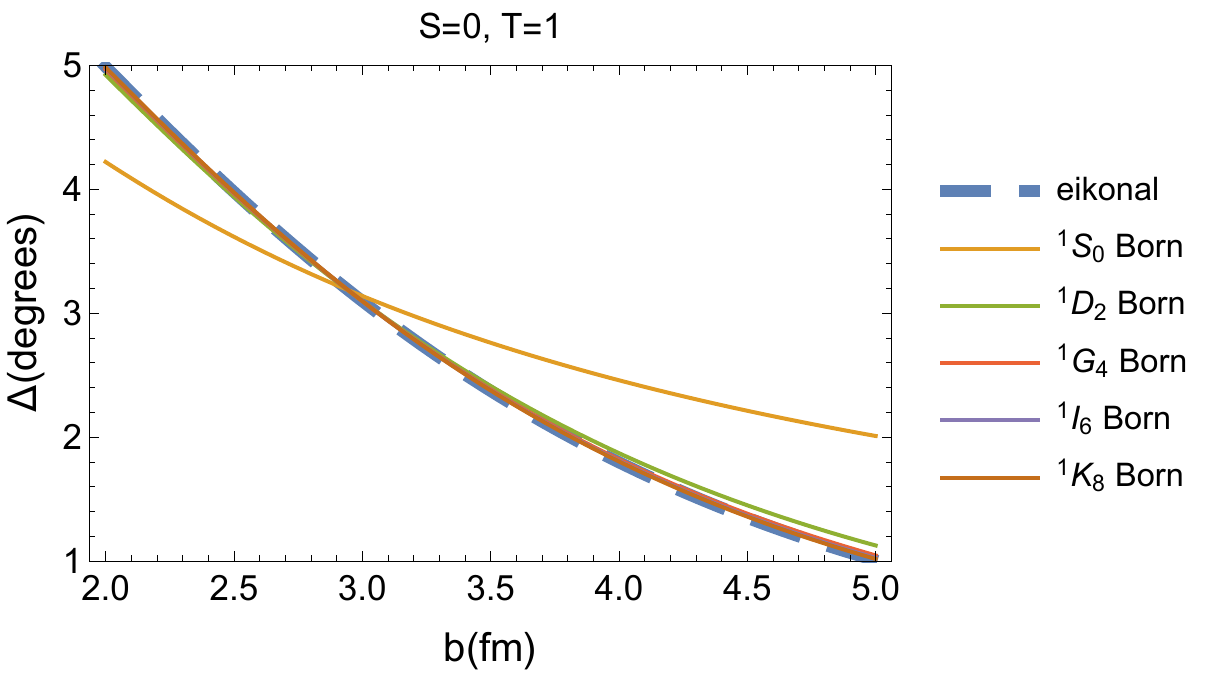}
\includegraphics[width=8cm]{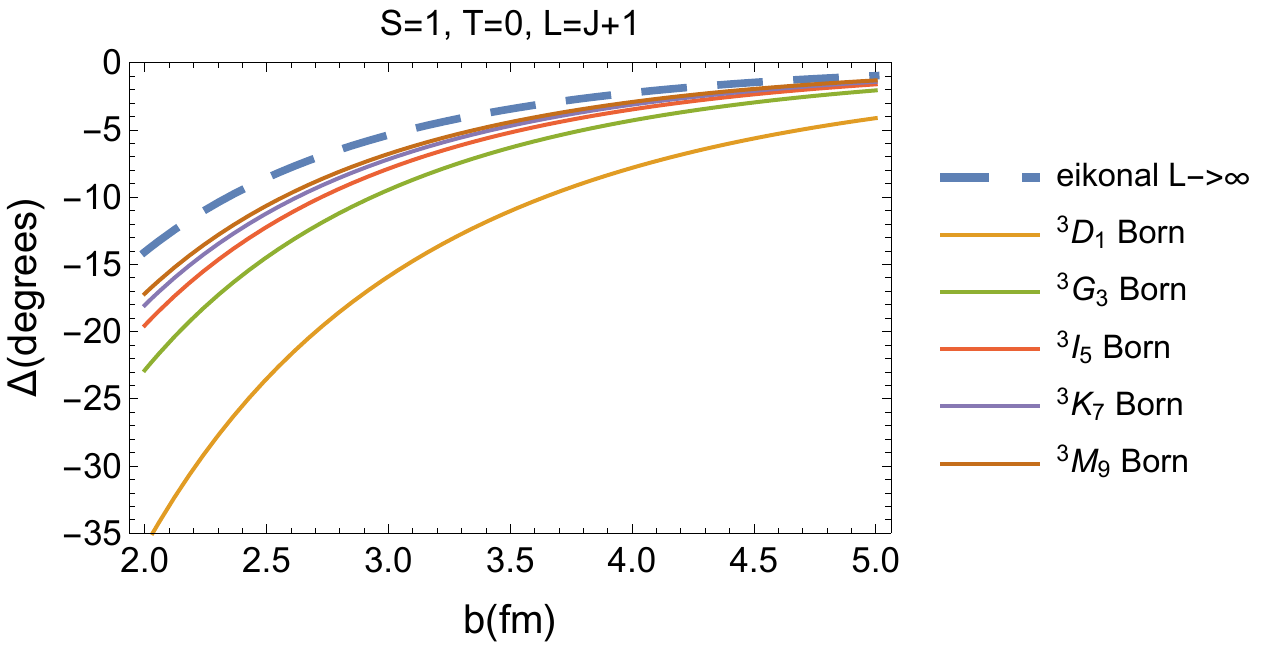}
\includegraphics[width=8cm]{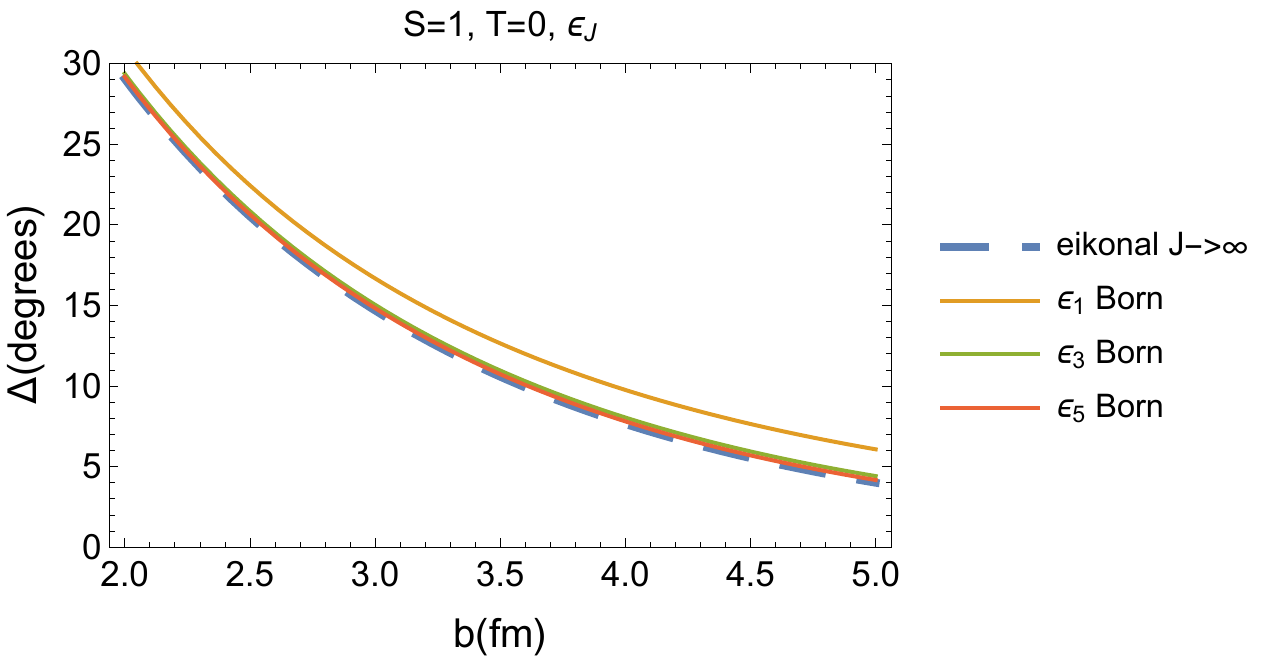}
\includegraphics[width=8cm]{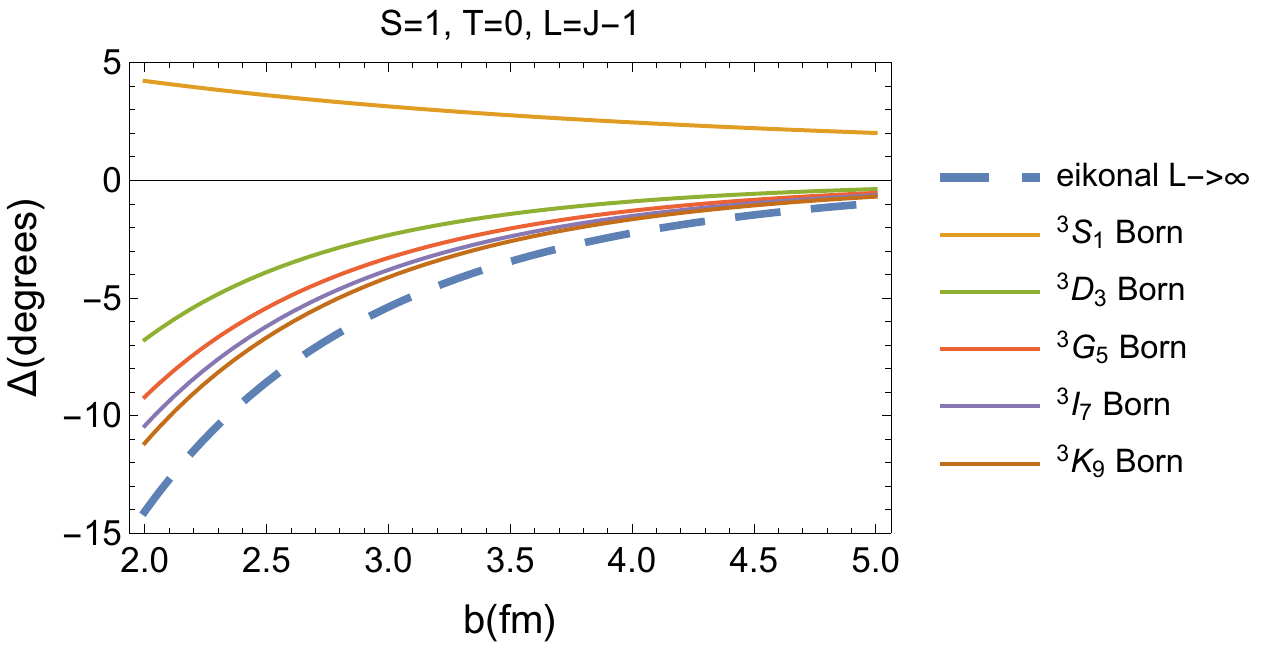}
\includegraphics[width=8cm]{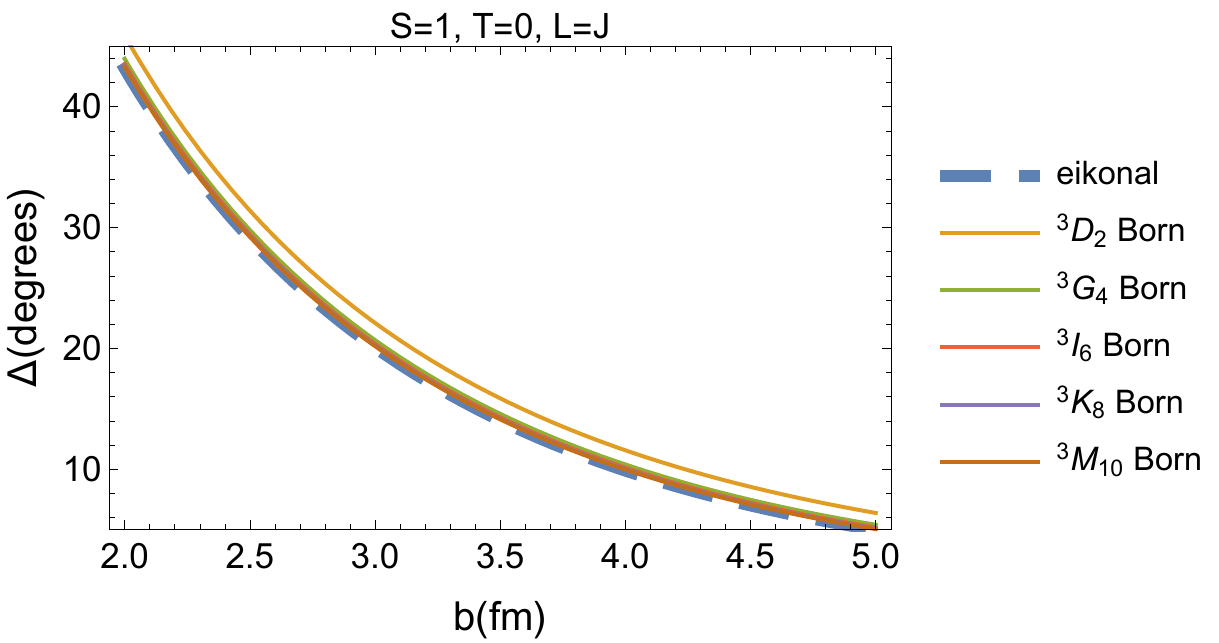}
\includegraphics[width=8cm]{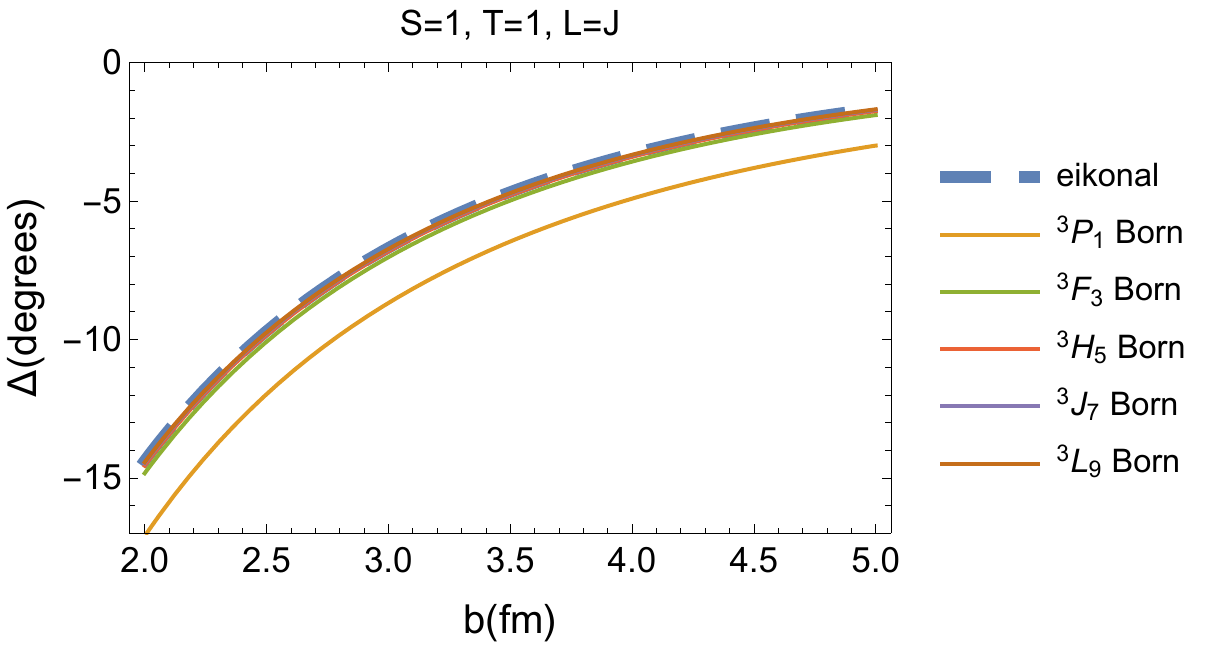}
\includegraphics[width=8cm]{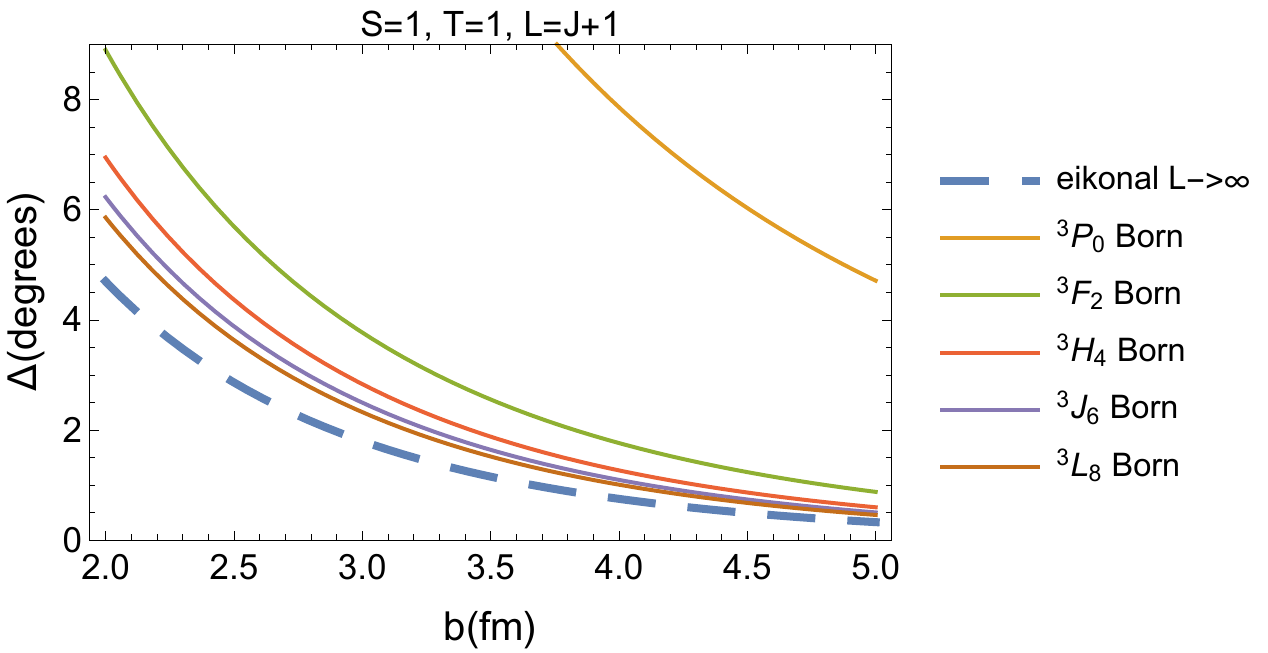}
\includegraphics[width=8cm]{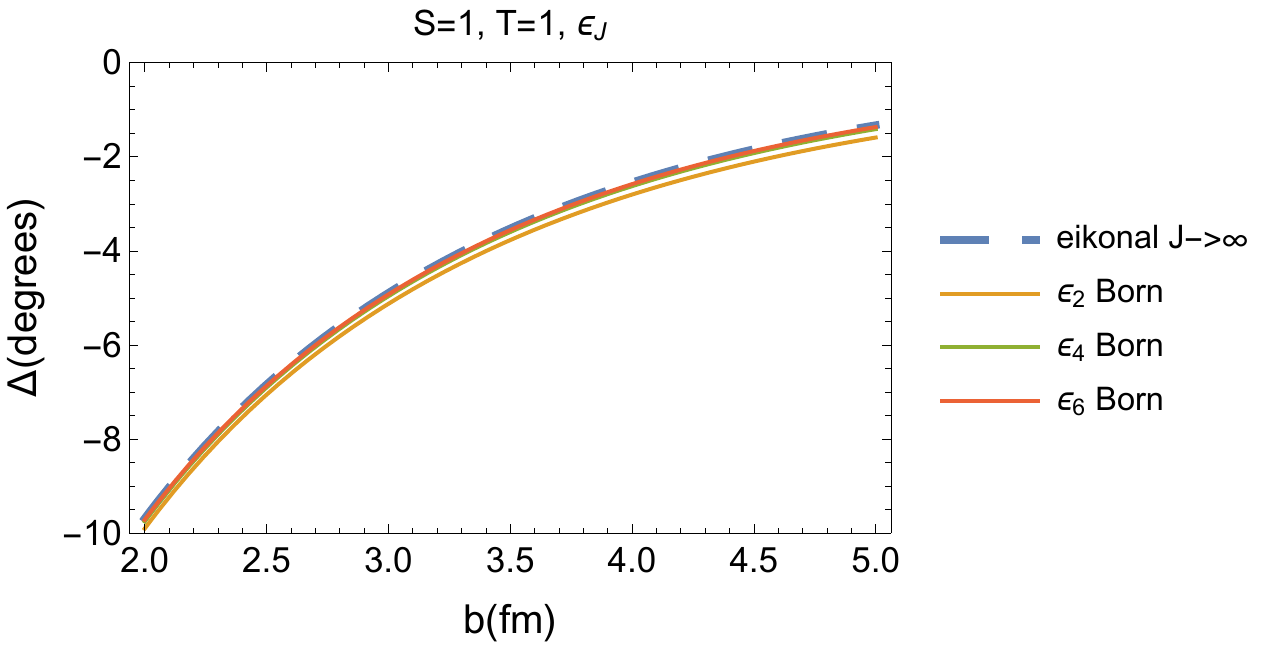}
\includegraphics[width=8cm]{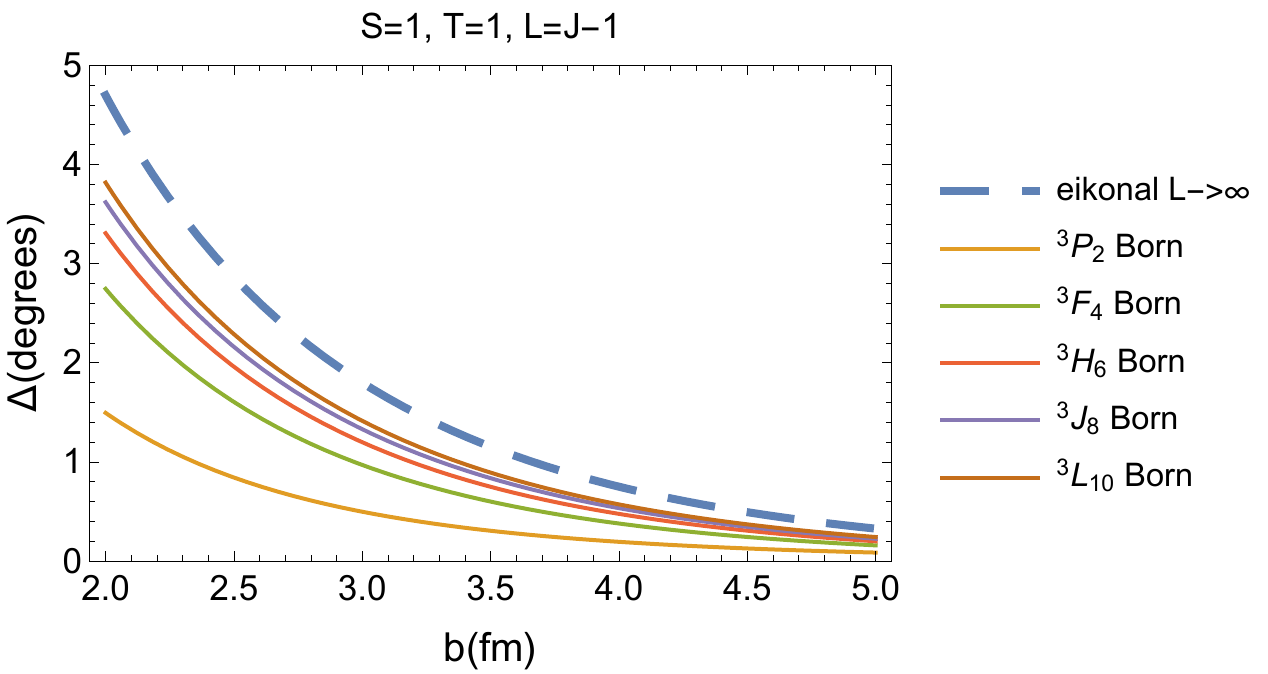}
\caption{Scaled phase-shifts $\Delta_{SLJ} $ for the
  OPE-potential computed  in  Born and  eikonal approximations
  as a function of the impact parameter $b$.}
\label{fig:born}      
\end{figure*}

\section{The peripheral plot}
\label{sec:peripheral}

Our present motivation for an universal plot where many partial waves
share their long distance content comes from the eikonal
approximation. Let us consider the scattering by a {\it local}
potential $V(r)$ independent on the angular momentum\footnote{ For the
  long distance component, the locality and angular momentum
  assumptions are supported by the particle exchange picture as well
  as data analyses based on the $L$-dependent
  potentials~\cite{Perez:2014yla,Perez:2014waa}, where terms $L^2$ are
  non-vanishing only at short distances.}.  The phase-shifts in the
eikonal approximation are given by
\begin{eqnarray}
\delta_L (p) = - \frac{M}{2p} \int_b^\infty dr \frac{r}{\sqrt{r^2-b^2}} V(r) \, . 
\label{eq:eikonal} 
\end{eqnarray}
This result also holds in the WKB approximation for a potential much
smaller than the centrifugal barrier~\cite{pascualquantum}.  In this
expression the impact parameter acts as a short distance
regulator. The interesting feature of the eikonal approximation is
that this property remains true to all orders in the eikonal
expansion~\cite{Wallace:1973iu}. This is unlike standard perturbation
theory based on expansions in powers of the potential, where this
contributes at all points, i.e. $0 \le r < \infty$, and this becomes
critical for singular potentials, like OPE (specially the tensor
force) or chiral TPE potentials\footnote{In fact, it was shown in
  appendix C of Ref. \cite{PavonValderrama:2005uj} that for potentials
  more singular than $\frac{1}{r^2}$ at the origin, there is always a
  finite order in standard perturbation theory where the lowest-lying
  partial waves give divergent contributions for the phase-shifts,
  even although the Born approximation was finite for these partial
  waves.}  (see \cite{PavonValderrama:2005wv,
  PavonValderrama:2005uj}).

We define the scaled phase-shift as
\begin{eqnarray}
\Delta_L (b) \equiv \left(L+\frac12 \right) \delta_L (p) \, . 
\label{delta} 
\end{eqnarray}
Using
Eq.~(\ref{eq:impact}) and Eq.~(\ref{eq:eikonal}), 
in the eikonal approximation we get
\begin{eqnarray}
\Delta_L (b) =
 - \frac{M b}{2} \int_b^\infty \, dr \, \frac{r}{\sqrt{r^2-b^2}} V(r)   \, . 
\label{eq:ikonal} 
\end{eqnarray}
Note that according to the eikonal approximation the scaled
phase-shift, $\Delta_L (b)$, {\it only} depends on the impact
parameter if the potential $V(r)$ does not depend on the angular
momentum. One expects to observe this scaling in the so-called {\it
  peripheral plot}, where $\Delta_L (b)$ is represented instead of
$\delta_L (p)$.  In the real situation encountered in a PWA we expect
violations to this scaling behavior only for low partial waves or for
small impact parameters. In general, the explicit angular momentum
dependence of the tensor interaction induces the most important
violation. Nonetheless, in this work we will see that in several cases
these relations work rather accurately within the experimental
uncertainties.

While the discussion on peripherality for a central potential is
rather straightforward, complications arise in presence of the NN
tensor force.  In particular, a direct eikonal three-dimensional
treatment of the tensor force~\cite{besprosvany1997eikonal}, unlike
the central force case, generates ambiguities. The alternative
discussion based on a WKB approach to coupled channel partial
waves~\cite{christian1950neutron} provides no way to calculate the
mixing parameters $\epsilon_{SLJ}$.  In this work we introduce an
alternative approach by using first-order perturbation theory, and
then taking the eikonal limit according to the WKB approximation.

Results are particularly simple for the nuclear bar representation and
we refer to Appendices \ref{sec:pert-ope} and \ref{sec:WKB} for
details.  In first-order perturbation theory we have (see
Appendix~\ref{sec:pert-ope})
\begin{eqnarray}
\bar \delta_{SJJ} (p) &=& - \frac{M}{p} \int_0^\infty dr \, \left[ \hat j_{J}(pr) \right]^2 V_{J,J}^J (r) \, , \label{eq:deltabar_l_j}
\\
\bar \delta_{SLJ}^\pm(p) &=& - \frac{M}{p} \int_0^\infty dr \,  \left[ \hat j_{J \pm 1}(pr)\right]^2 V_{J\pm 1,J\pm 1}^J(r)  \, , \label{deltabar_pm}
\\
\bar \epsilon_{SLJ}(p) &=& - \frac{M}{p} \int_0^\infty  dr \, \hat j_{J-1}(pr) \hat{j}_{J+1} (pr) V_{J-1,J+1}^J (r) \,  
\label{eq:ps-OPE}
\end{eqnarray}
where $\hat j_l(x) \equiv x j_l(x)$ are the reduced spherical Bessel
functions.  Note that, in the above expressions, the potential
contributes everywhere. The $1/r^3$ singularity could be tamed by
introducing a short distance cut-off $r_c$, generating a cut-off
dependence which, however, becomes mild\footnote{ However, as shown in
  Ref.~\cite{PavonValderrama:2005uj} and already commented in the
  previous footnote, higher-order perturbation theory will reinstate
  the short distance divergence.  } for $J>2$.

We now define the scaled phase-shifts for $S=1$,  
\begin{eqnarray}
\Delta_{SLJ}^- &=& (J-1/2) \bar \delta_{SLJ}^- \label{deltam} \, , \\
\Delta_{SLJ}^\epsilon &=& (J+1/2) \bar \epsilon_{SLJ} \label{deltae} \, , \\
\Delta_{SLJ}^+ &=& (J+3/2) \bar \delta_{SLJ}^+ \label{deltap} \, ,  
\end{eqnarray}
in terms of the nuclear bar representation. We then apply the WKB
approximation, as shown in Appendix~\ref{sec:WKB}, obtaining very
simple results.  For the uncoupled channels Eq.~(\ref{eq:ikonal})
remains valid. In the coupled channels case, Eq.~(\ref{eq:ikonal})
also holds for the diagonal matrix element, so that the only
modification comes from the off-diagonal element,
\begin{eqnarray}
\Delta_{SLJ}^\pm (b)_{\rm WKB} = - \frac{M b}{2} \int_b^\infty dr \, 
\frac{r}{\sqrt{r^2-b^2}} \, V^J_{J\pm1,J\pm 1} (r) \, ,  
\label{deltapmWKB}
\\ 
\Delta_{SLJ}^\epsilon(b)_{\rm WKB}= 
- \frac{M b}{2}\int_b^\infty dr \,  \frac{2 b^2-r^2}{r
  \sqrt{r^2-b^2}} \, V_{J-1,J+1}^J (r) \, . 
\label{deltaeWKB}
\end{eqnarray}
In Fig.~\ref{fig:born} we compare the values of the $\Delta$'s both
for the eikonal approximation and the perturbative OPE result. As we
see, the scaling holds already for moderate values of the angular
momentum $J$, especially for the uncoupled partial waves and the
mixing phases $\epsilon_J$. In the eikonal calculations shown in
Fig. \ref{fig:born} for the coupled channels, the limit $L\rightarrow
\infty$ has been taken in the matrix elements of the OPE potential
appearing in Eqs. (\ref{deltapmWKB}) and (\ref{deltaeWKB}). This is
important to get rid of the explicit $L$-dependence in these matrix
elements, due to the presence of the tensor operator $S_{12}$.

 It is also worth noting that the convergence of the scaled
 phase-shifts for the coupled partial waves computed in the Born
 approximation with the OPE potential is much slower than in the other
 waves, as shown in Fig. \ref{fig:born}. This effect is, again, due to
 the explicit $L$-dependence of the potential matrix elements in these
 channels. However, we have checked that increasing the angular
 momentum $L$, the Born OPE results eventually reach the eikonal
 curves indicated by the dashed lines in Fig. \ref{fig:born}.

\begin{figure*}
\centering
\epsfig{file= 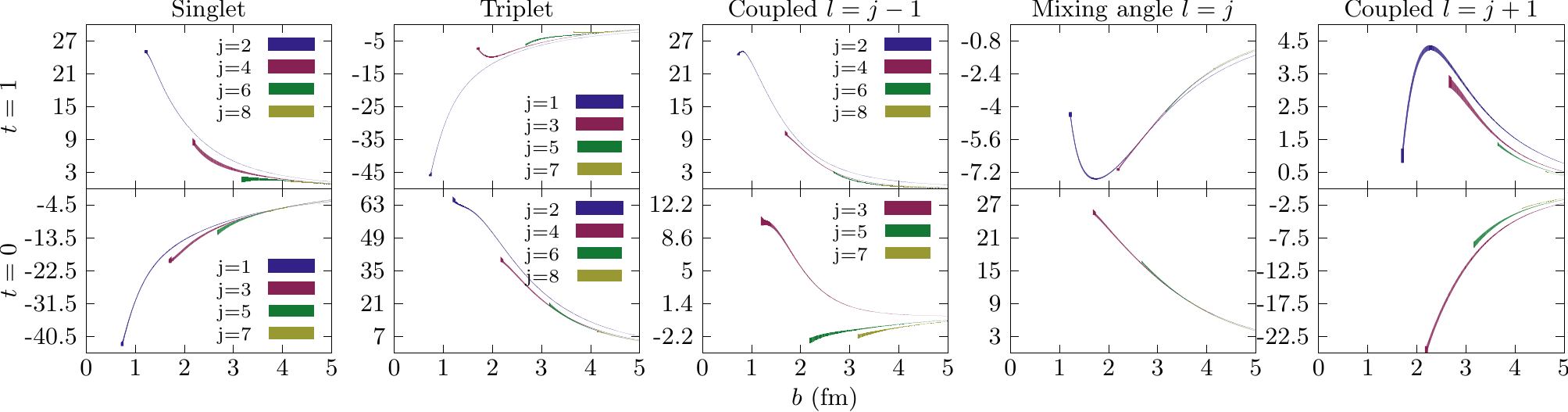,width=18cm}
\caption{Peripheral plots for the scaled phase-shifts
(in degrees) $(L + 1/2)\delta_{SLJ} (p)$ 
  with $L=J, J\pm 1$ and $p b = (L+1/2)$ and mixing
  angles $(J + 1/2) \epsilon_{J} (p)$ with $p b = (J+1/2)$ 
   as a function of the impact parameter $b$ (in fm).  The
  statistical errors for the DS-OPE
  PWA are taken from Refs ~\cite{Perez:2013mwa,Perez:2013jpa}. }
\label{fig:eik-stat}       
\end{figure*}
\begin{figure*}
\centering
\epsfig{file=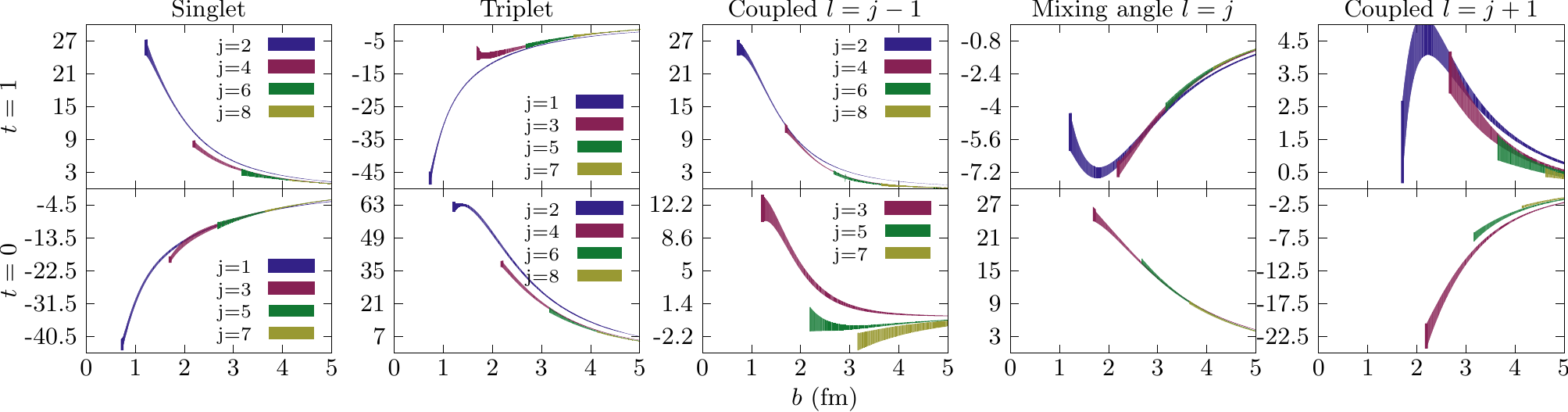,width=18cm}
\caption{Same as Fig.~\ref{fig:eik-stat} but for phase-shifts coming 
from 13 high-quality fits in all partial waves with $J \le 5$, as a function
  of the impact parameter. We show the Nijmegen 
 PWA~\cite{Stoks:1993tb}, Nijm I, NijmII,
  Reid93~\cite{Stoks:1994wp}, the AV18~\cite{Wiringa:1994wb}, CD
  Bonn~\cite{Machleidt:2000ge}, Spectator~\cite{Gross:2008ps}
  and the six Granada potentials denoted 
  as DS-OPE~\cite{Perez:2013mwa,Perez:2013jpa},
  DS-$\chi$TPE~\cite{Perez:2013oba,Perez:2013cza},
  SOG-OPE~\cite{Perez:2014yla}, SOG-$\chi$TPE, DS-$\Delta$BO and
  SOG-$\Delta$BO~\cite{Perez:2014waa}.}
\label{fig:eik-sys}       
\end{figure*}

\section{Peripheral plots for high quality interactions}
\label{sec:peripheral-hq}

As we have already mentioned, we seek for a more direct relation
between the configuration space extension of the potential and the
determined phase-shifts. Motivated by the perturbative analysis of the
previous section, we show in Fig.~\ref{fig:eik-stat} the scaled
phase-shifts for the DS-OPE potential obtained in the Granada-2013 PWA
\cite{Perez:2013mwa,Perez:2013jpa}.  All different channels with $J
\le 6$ are shown grouped among uncoupled (singlet and triplet panels)
and coupled channels. The three scaled phase-shifts ($\Delta^-,
\Delta^\epsilon, \Delta^+$) for coupled channels were defined in
Eqs. (\ref{deltam}--\ref{deltap}) and in the figure are labeled with
the effective value of the angular momentum applied in the peripheral
plot, which is $J-1, J$ and $J+1$, respectively.  In the case of the
uncoupled channels the value of the angular momentum $L=J$ is used to
scale the phase-shifts as defined in Eq. (\ref{delta}).

According to our previous discussion, impact parameters below the
elementariness radius $b < r'_c = 1.8$ fm most likely probe nucleon
finite size effects and hence we expect the largest deviations in this
range, as it can be checked from Fig.~\ref{fig:eik-stat}.  The level
of scaling is remarkable in the uncoupled channels, where the scaled
phase-shifts almost overlap for $L>2$.  In the coupled channels, the
scaling violations for the $\Delta^\pm$ have to do with the explicit
$L$-dependence of the diagonal components of the potential.  The
off-diagonal potential has a weaker $L$-dependence, resulting in a
remarkable degree of scaling of the $\Delta^\epsilon$.  The panels of
Fig. \ref{fig:eik-stat} share the same similarities already observed
in those of Fig. \ref{fig:born}, thus resembling the OPE perturbative
treatment of the previous section, but going beyond it by considering
all orders.

In Fig. ~\ref{fig:eik-stat} we also display the corresponding
statistical error bars. As we see these errors are tiny in the
peripheral plot (they are even smaller than a typical line-width). As
anticipated in our discussion above, the reason of these very small
errors is that, once the $f^2$ coupling constant is fixed to large
accuracy, errors above $b> 1.8$ fm stem only from higher order
perturbative corrections beyond the Born approximation.  The
usefulness of this figure is that, besides packing all phase-shifts
for fixed isospin into five separate groups with similar behavior, for
data below pion production threshold the range of the potential being
probed becomes rather obvious in terms of the impact parameter.

It is also interesting to compare the peripheral plot for some of the
available high-quality interactions sharing the {\it same} CD-OPE
potential with the {\it same} and common coupling constant
$f^2=0.075$\footnote{The most recent determination yields
  $f^2=0.0763(1)$~\cite{Perez:2016aol} but we prefer to keep the same
  old value to enhance the long distance equivalence of all
  potentials.}. They correspond to the Nijmegen PWA results denoted
Nijm I, NijmII, and Reid93~\cite{Stoks:1994wp}, the AV18
\cite{Wiringa:1994wb}, CD Bonn~\cite{Machleidt:2000ge}, and
Spectator~\cite{Gross:2008ps} potentials, and to the six Granada
potentials denoted DS-OPE, DS-$\chi$TPE, DS-$\Delta$BO, SOG-OPE,
SOG-$\chi$TPE and SOG-$\Delta$BO
\cite{Perez:2013oba,Perez:2013cza,Perez:2014yla}. To this end we
define the mean and the standard deviation as usual,
\begin{eqnarray}
  {\rm Mean}(\delta) &=& \frac1{N} \sum_{i=1}^N \delta^{i}
  \label{eq:mean} \, , \\ 
        {\rm Std}(\delta) &=& \sqrt{\frac1{N-1} \sum_{i=1}^N (\delta^{i} - {\rm Mean}(\delta))^2 } \, , 
        \label{eq:std}
\end{eqnarray}
and similarly for the scaled phase-shifts $\Delta$. 

In Fig.~\ref{fig:eik-sys} we show the spreading band of the scaled
phase-shifts generated by these different PWA, which describe their
contemporary NN scattering database.  As we have stressed in
Ref.~\cite{Perez:2014waa}, this can be considered as a lower bound on
the estimate of the systematic errors as a function of the
inter-nucleon distances being explored.  The larger errors are
concentrated at short distances, and this reflects our lack of
knowledge on how the interaction should be parameterized in the short
distance region.  While the true errors are expected to be larger, in
the long distance region they are very small due to the fixed long
distance behavior of the potential.

\begin{figure}
\centering
\includegraphics[height=5cm]{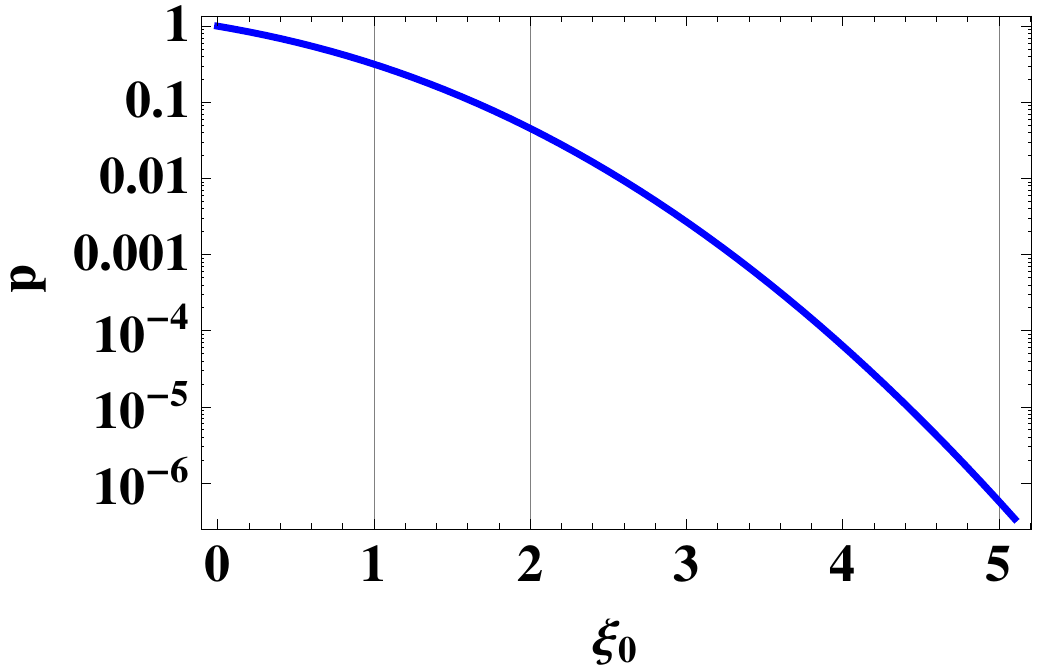}
\caption{The probability of {\em not being an outlier} as a function
  of the standardized discrepancy value $\xi_0$ (see main text). We
  mark the values corresponding to 1,2 and 5 standard deviations.}
\label{fig:p-value}       
\end{figure}

\begin{figure*}
\centering
\includegraphics[height=3.5cm]{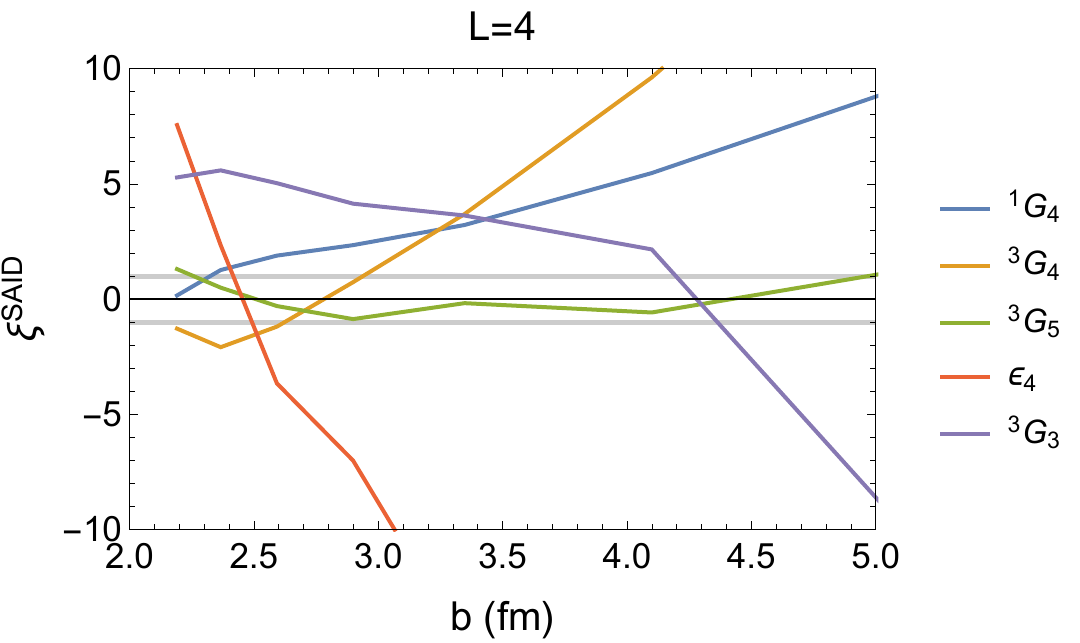}
\includegraphics[height=3.5cm]{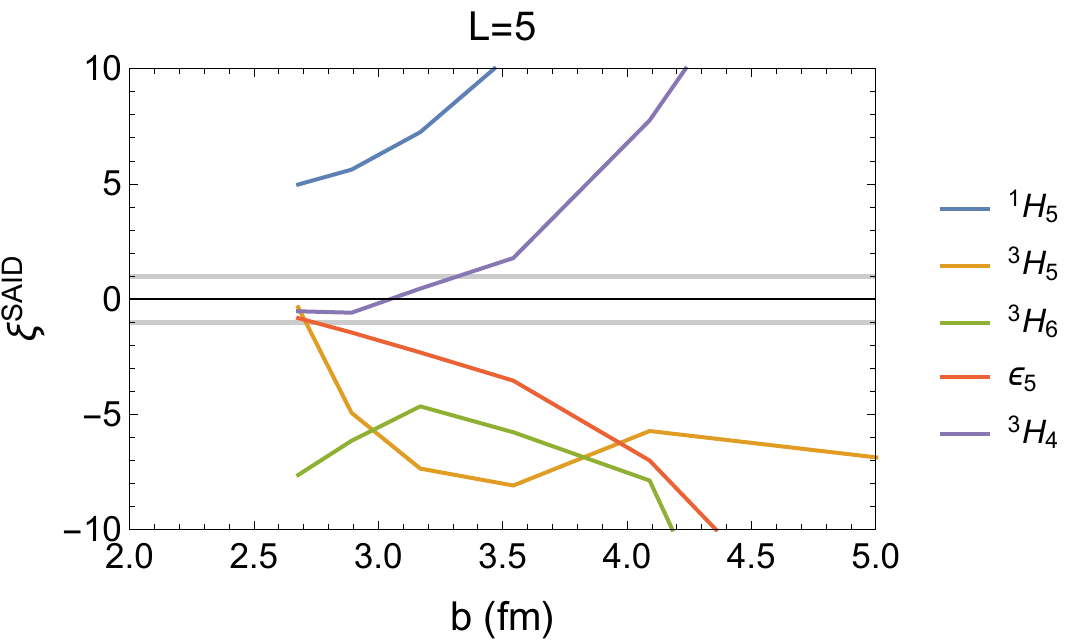}
\includegraphics[height=3.5cm]{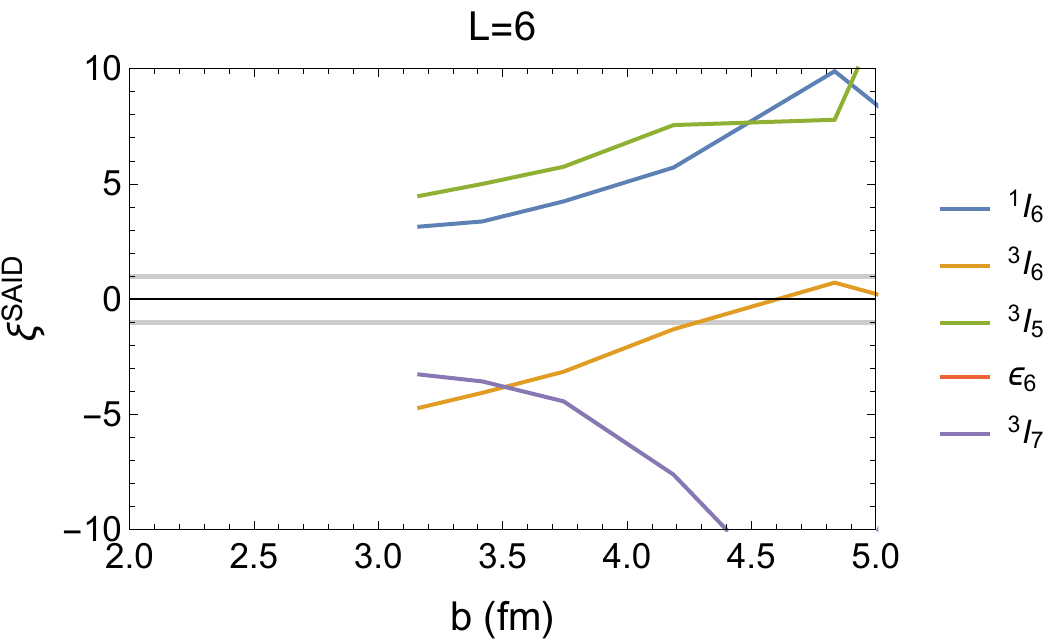} \\
\vskip.5cm 
\includegraphics[height=3.5cm]{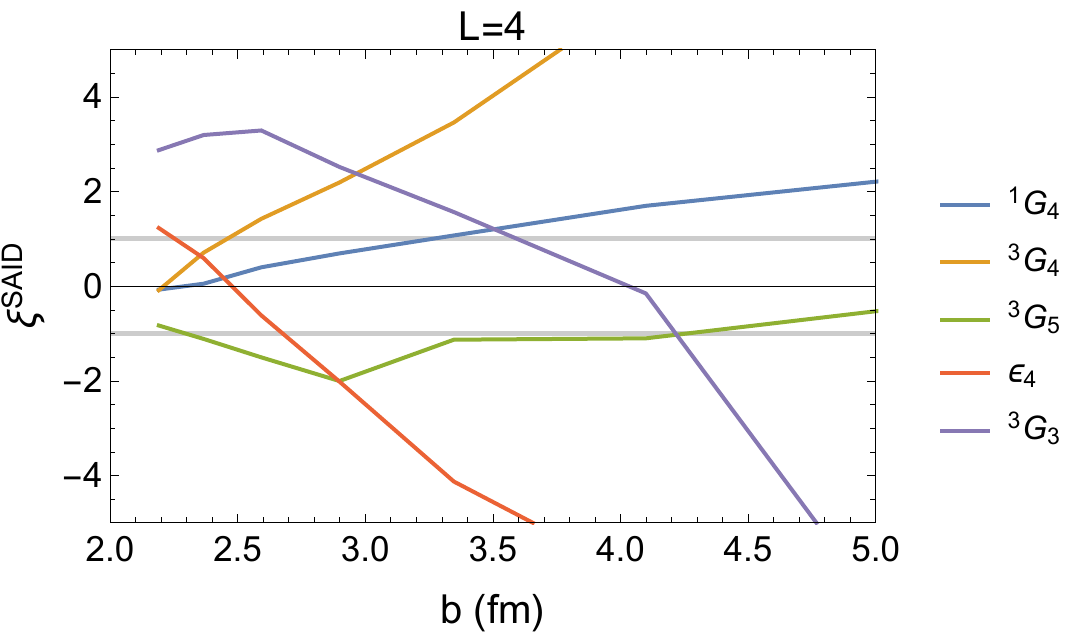}
\includegraphics[height=3.5cm]{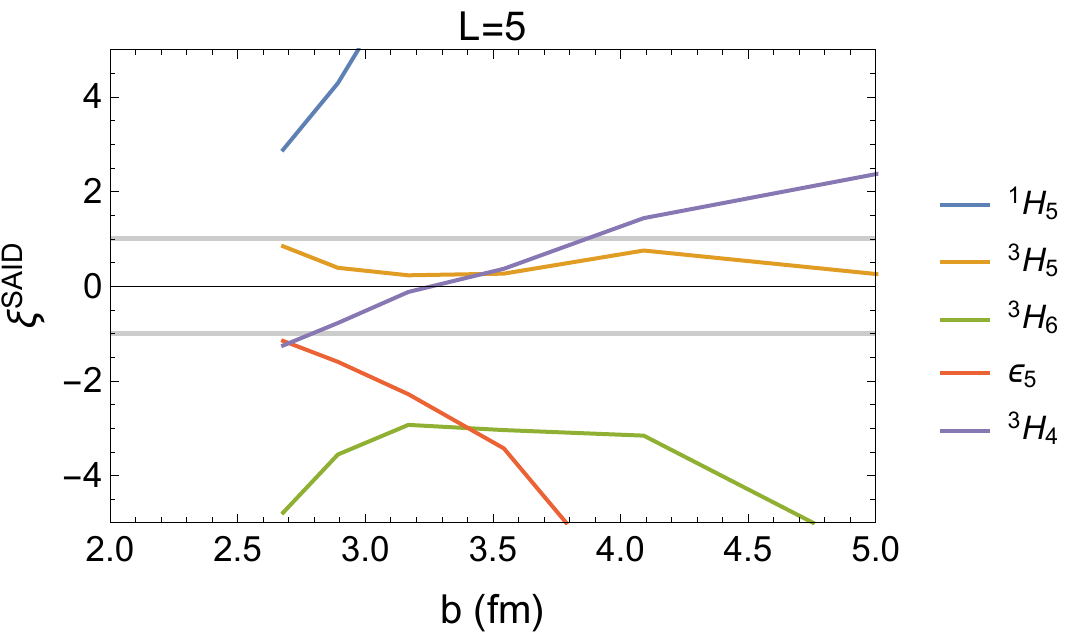}
\includegraphics[height=3.5cm]{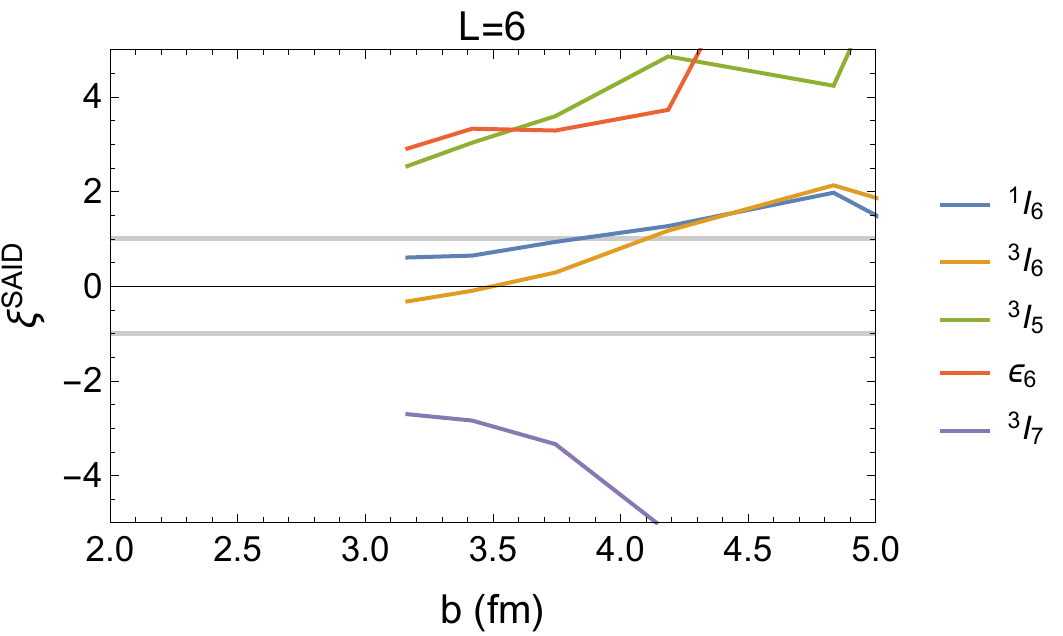} \\
\vskip.5cm 
\includegraphics[height=3.5cm]{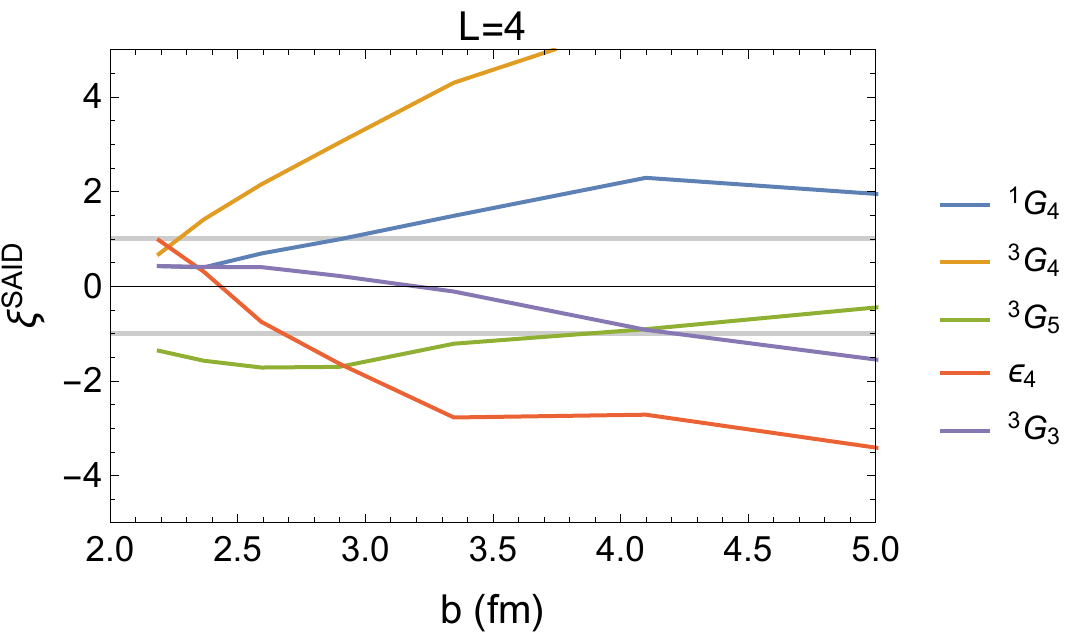}
\includegraphics[height=3.5cm]{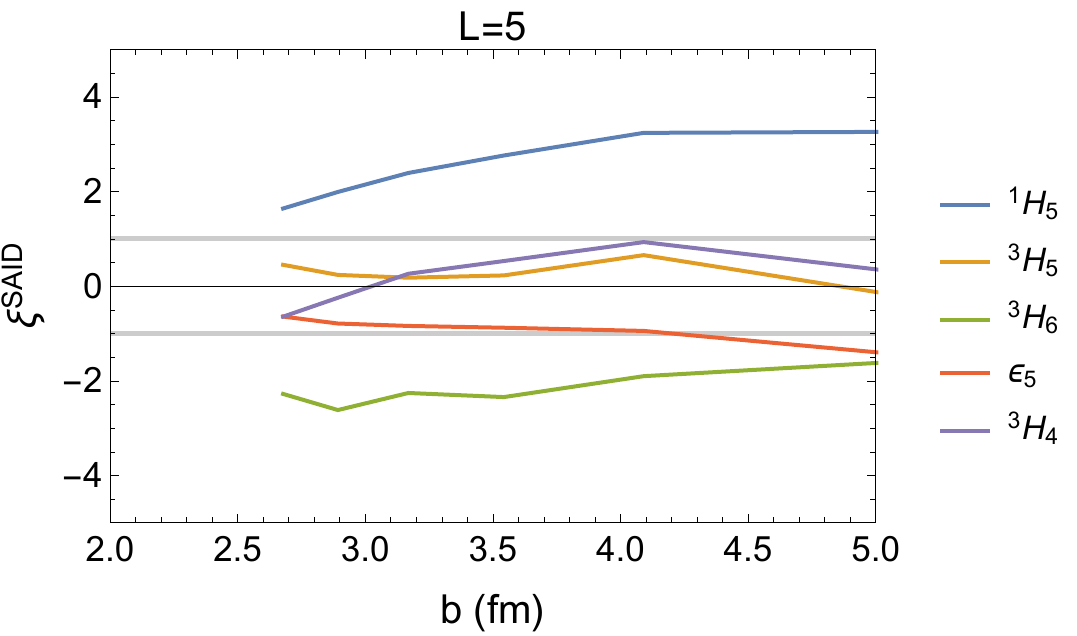}
\includegraphics[height=3.5cm]{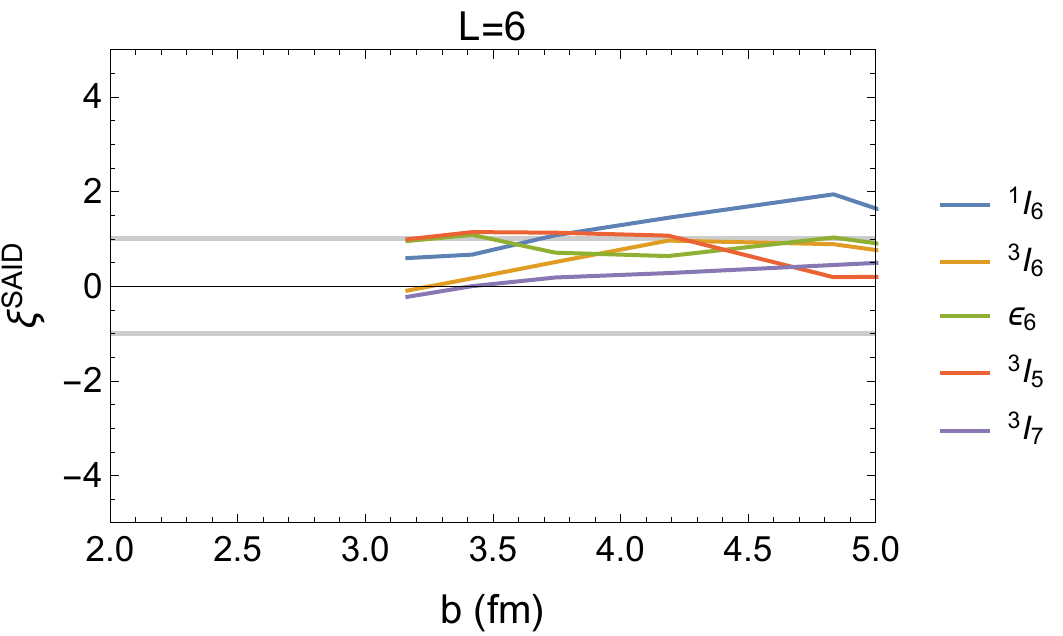} 
\caption{Comparison of the normalized discrepancy of the SAID NN
  solution (SM16)~\cite{Workman:2016ysf} with some sets of fits for $L=4$
  (left), $L=5$ (middle) and $L=6$ (right) as a function of the impact
  parameter $b=(L+1/2)/p$ for energies below $T_{\rm LAB} = 350$
  MeV. Top panels: comparison to the DS-OPE
  potential~\cite{Perez:2013mwa,Perez:2013jpa}.  Middle panels:
  comparison with the average of the 6 Granada-2013 potentials.
  Bottom panels: comparison with the average of the full set of 13
  high-quality potentials. Non-plotted discrepancies are out of
  range.}
\label{fig:xi-sys-SAID}       
\end{figure*}

\begin{figure*}
\centering
\includegraphics[height=3.5cm]{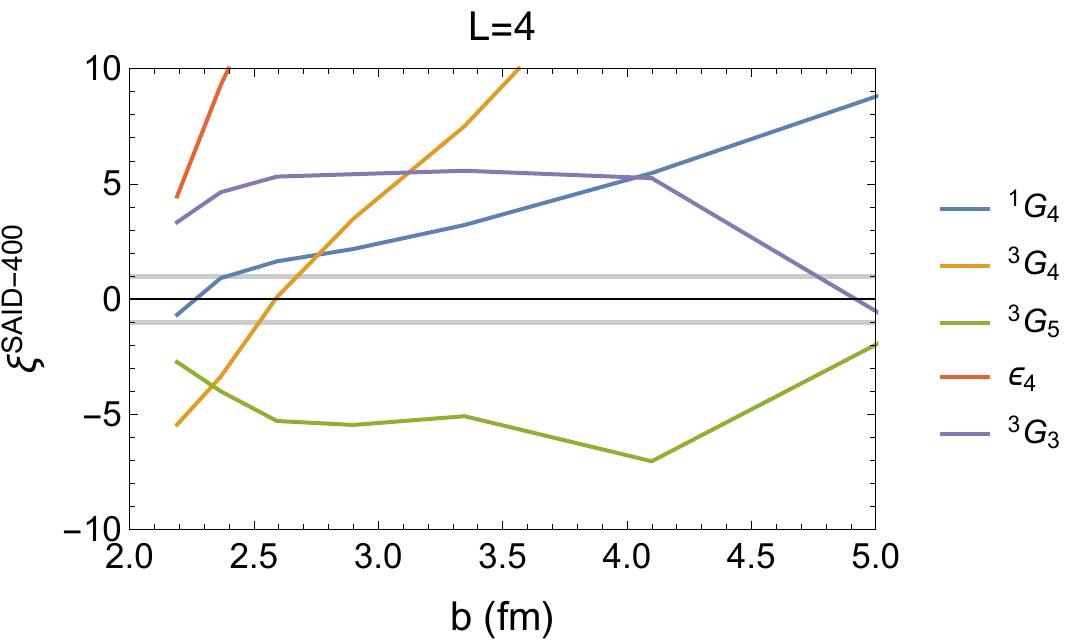}
\includegraphics[height=3.5cm]{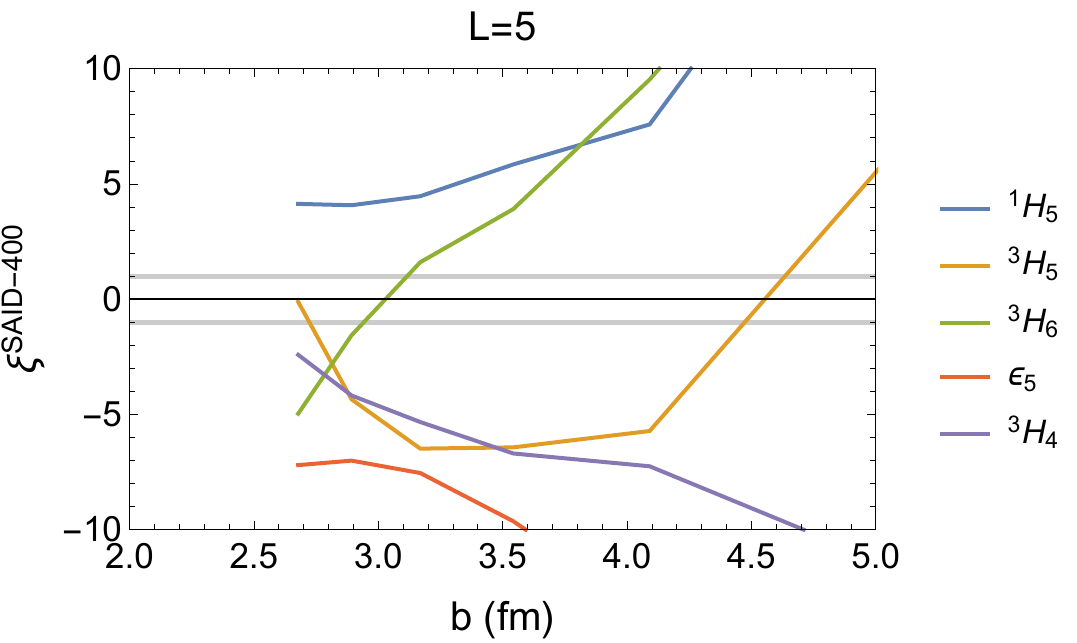}
\includegraphics[height=3.5cm]{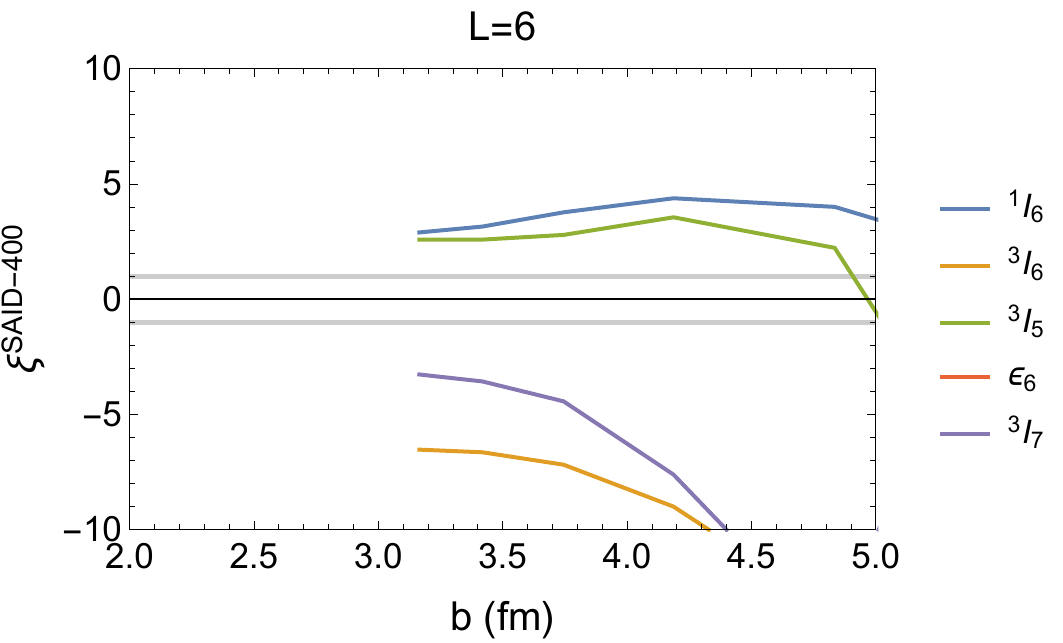} \\
\vskip.5cm 
\includegraphics[height=3.5cm]{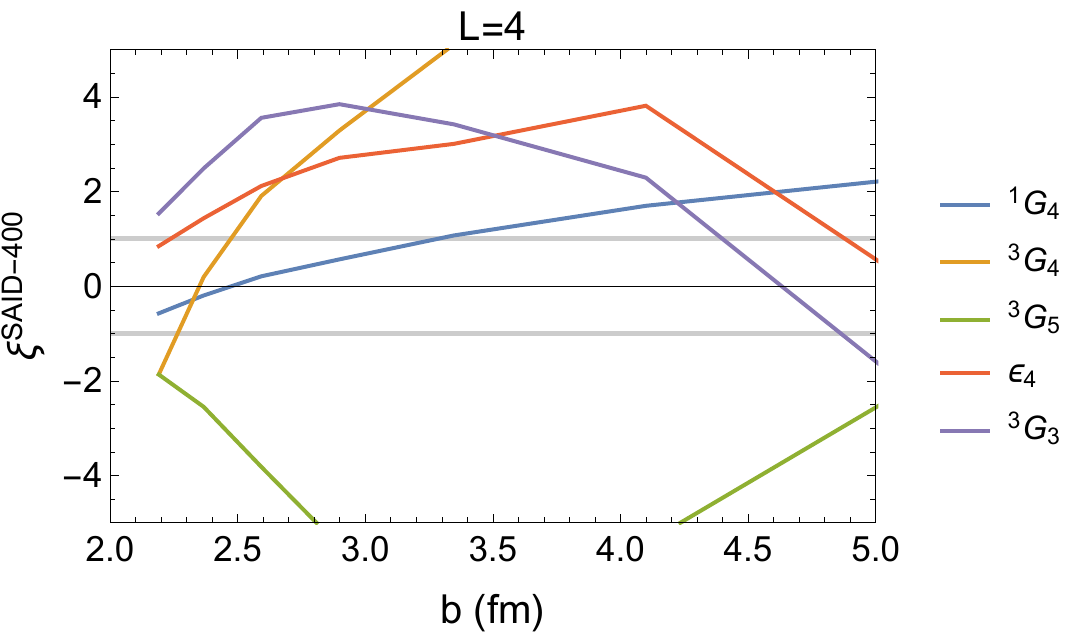}
\includegraphics[height=3.5cm]{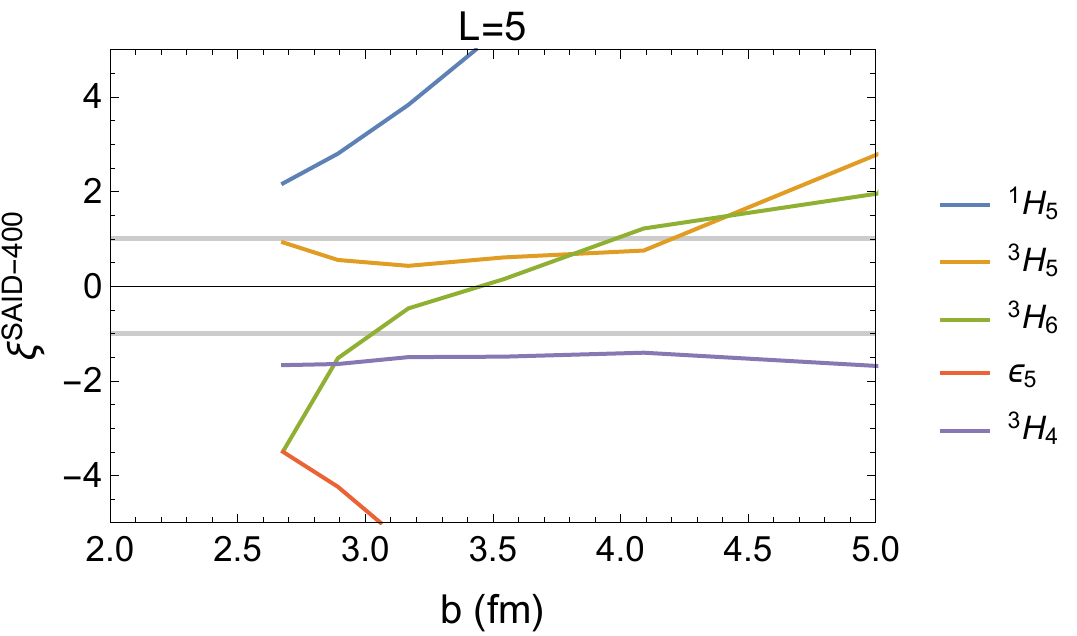}
\includegraphics[height=3.5cm]{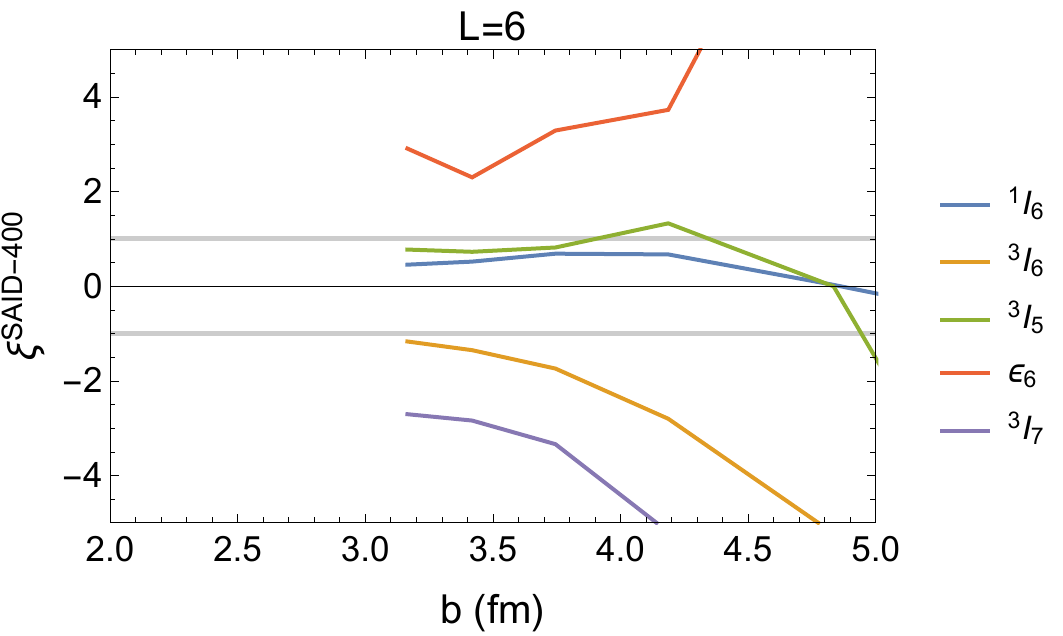} \\
\vskip.5cm 
\includegraphics[height=3.5cm]{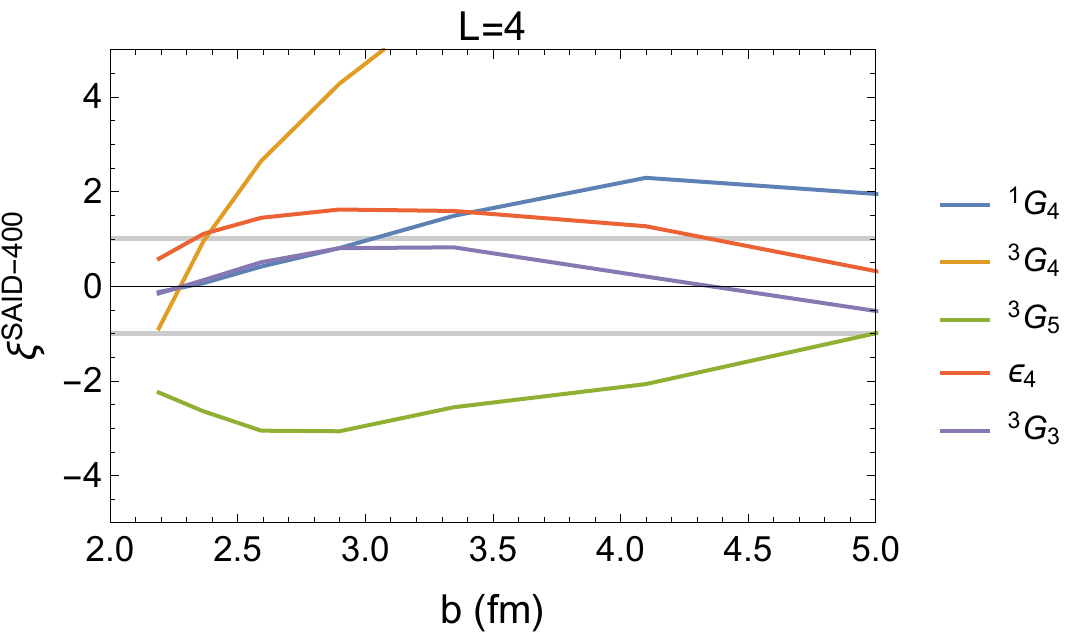}
\includegraphics[height=3.5cm]{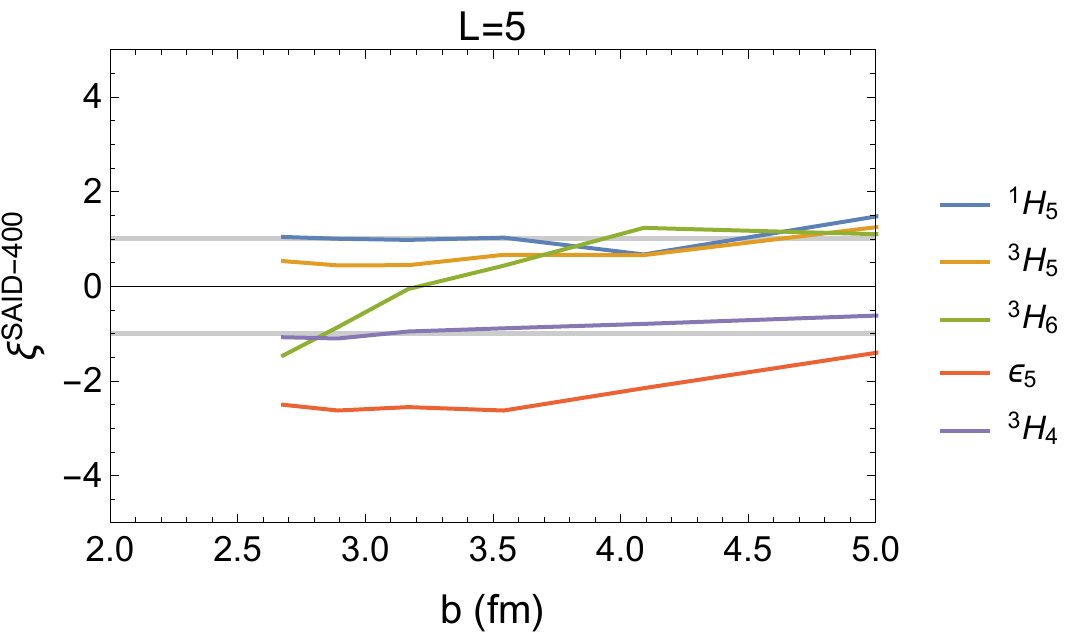}
\includegraphics[height=3.5cm]{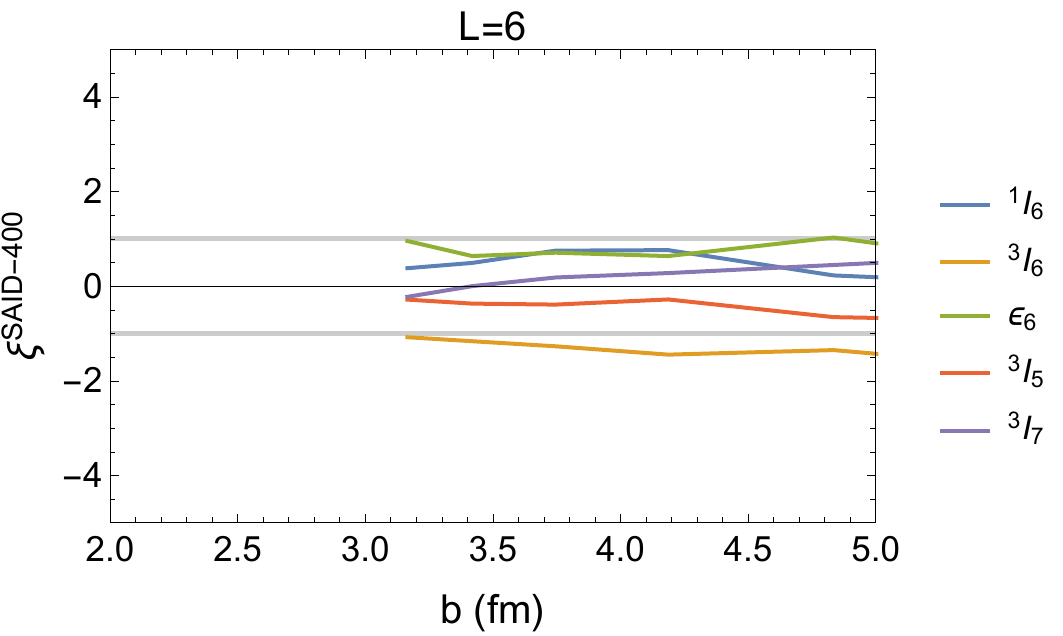} 
\caption{Same as Fig.~\ref{fig:xi-sys-SAID} for the SAID NN solution 
in the range $0-400$ MeV (SP40)~\cite{Arndt:2000xc}.}
\label{fig:xi-sys-SAID400}       
\end{figure*}

\begin{figure*}
\centering
\includegraphics[height=3.5cm]{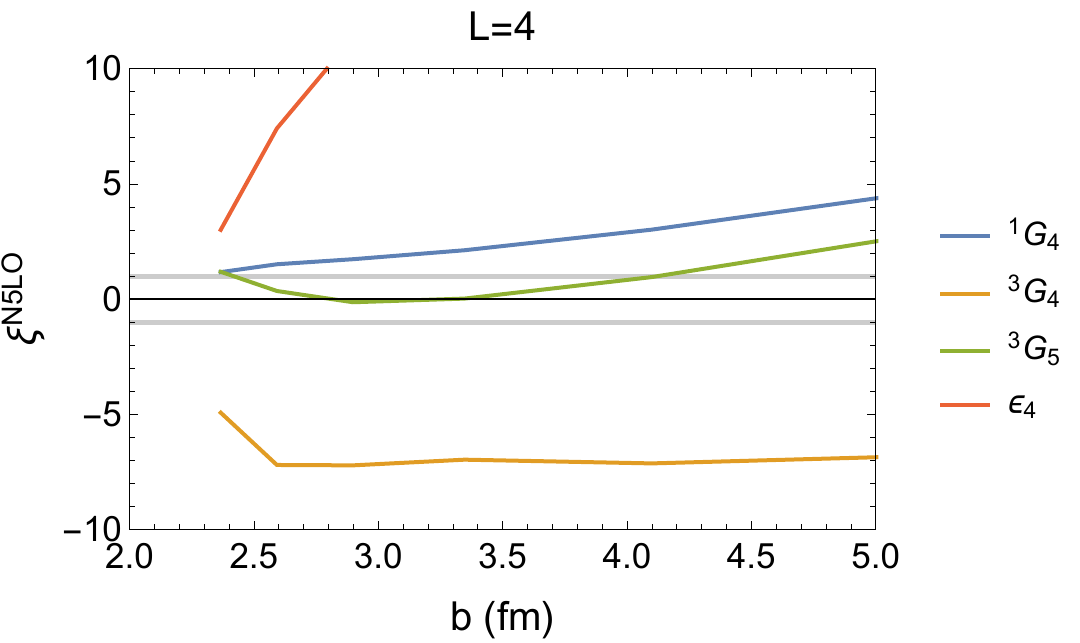}
\includegraphics[height=3.5cm]{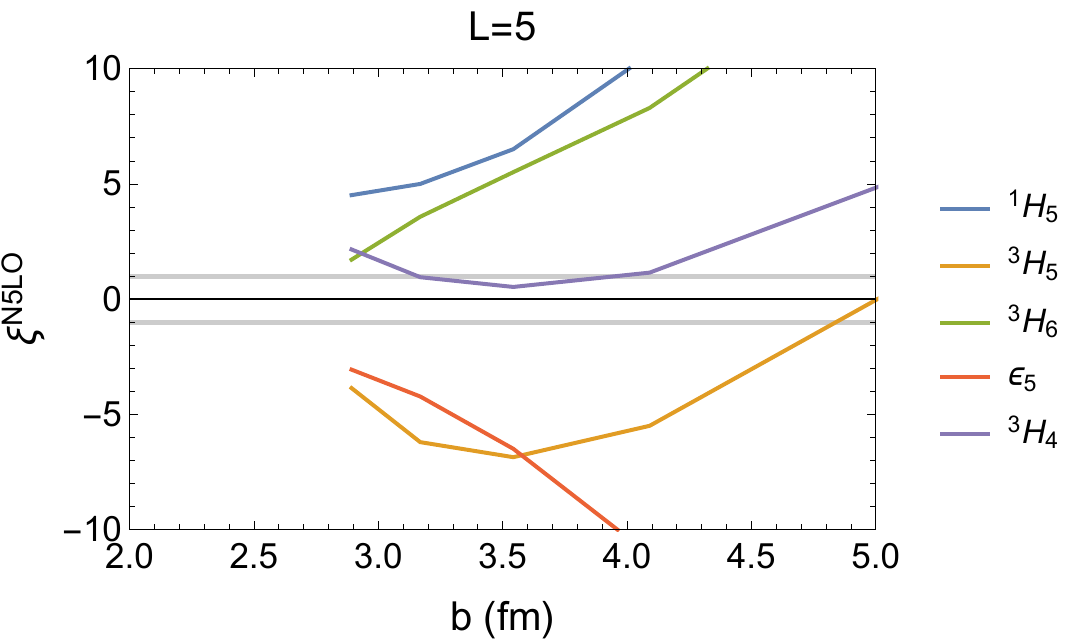}
\includegraphics[height=3.5cm]{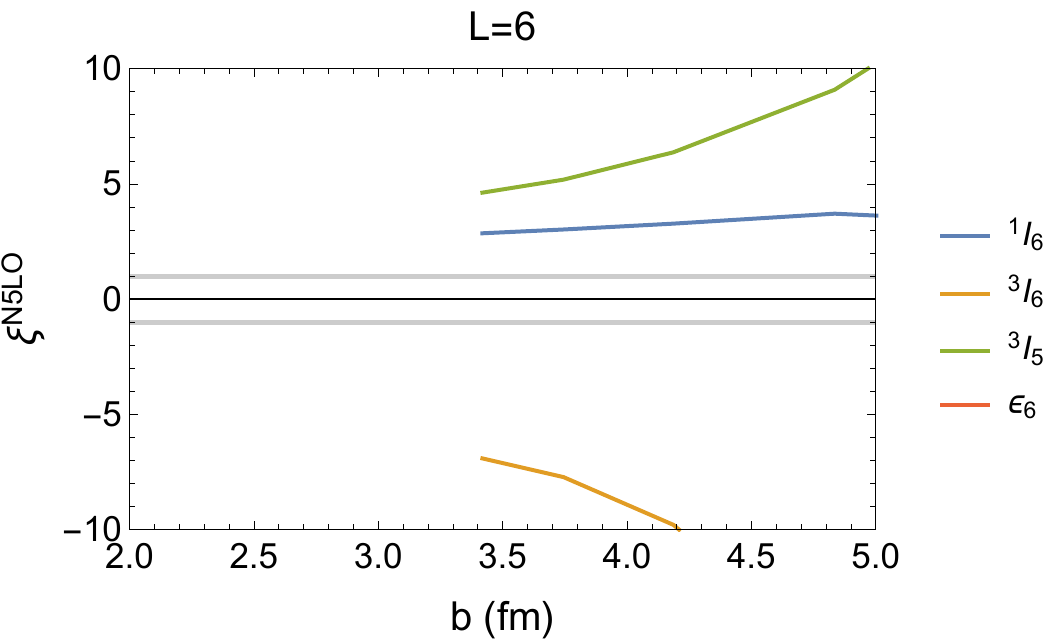} \\
\vskip.5cm 
\includegraphics[height=3.5cm]{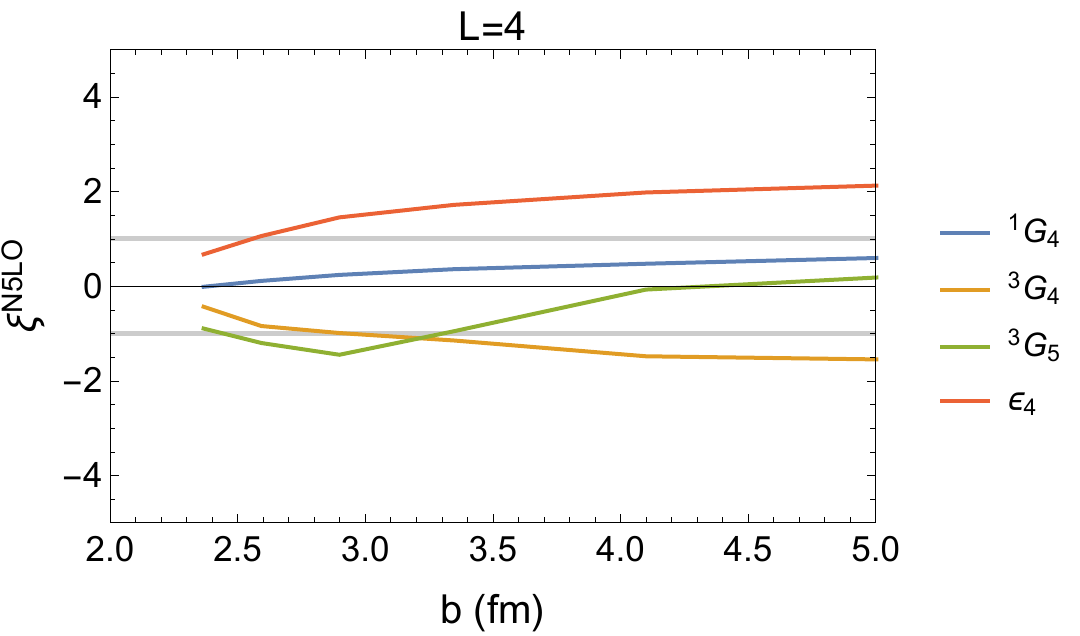}
\includegraphics[height=3.5cm]{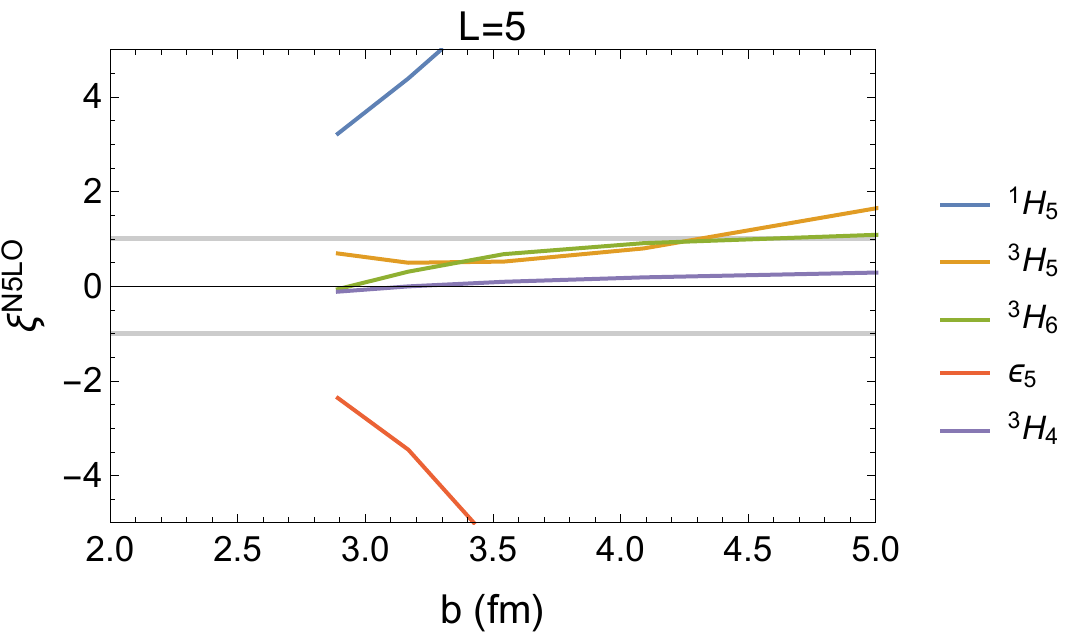}
\includegraphics[height=3.5cm]{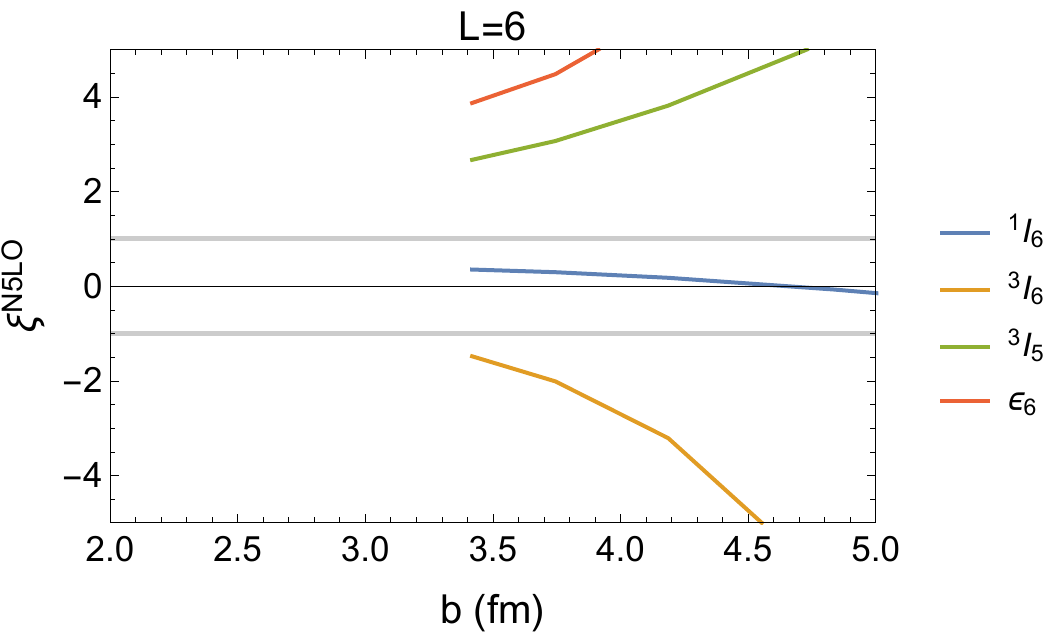} \\
\vskip.5cm 
\includegraphics[height=3.5cm]{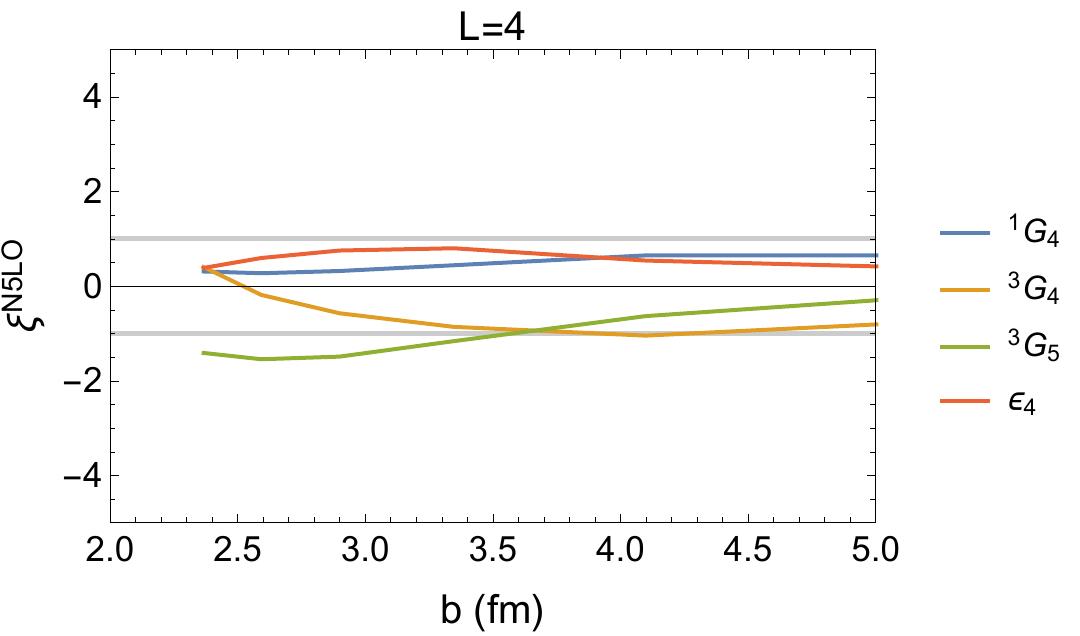}
\includegraphics[height=3.5cm]{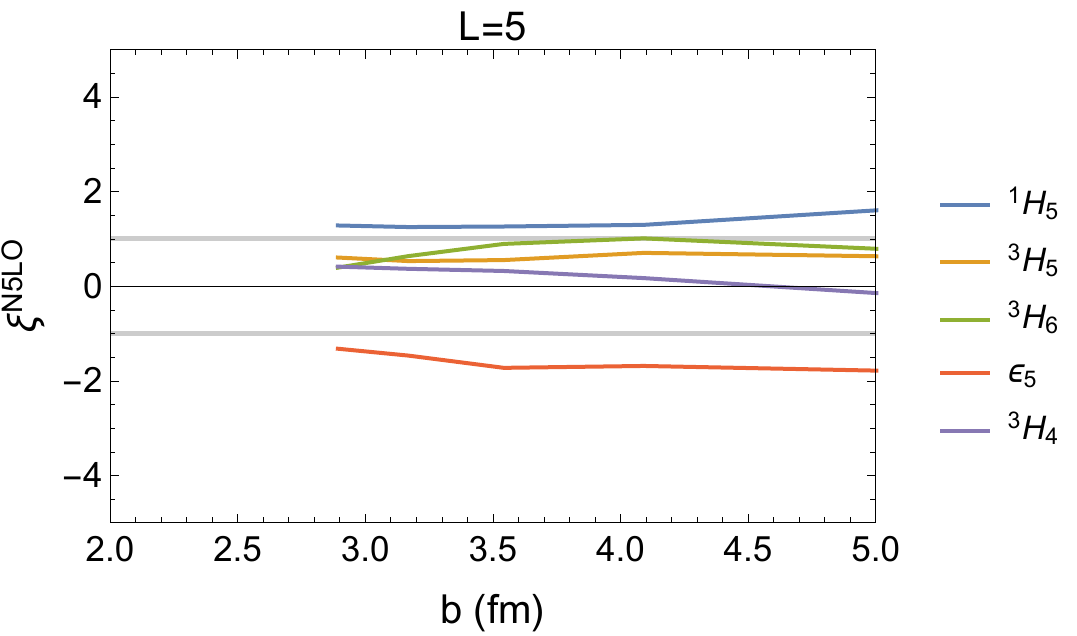}
\includegraphics[height=3.5cm]{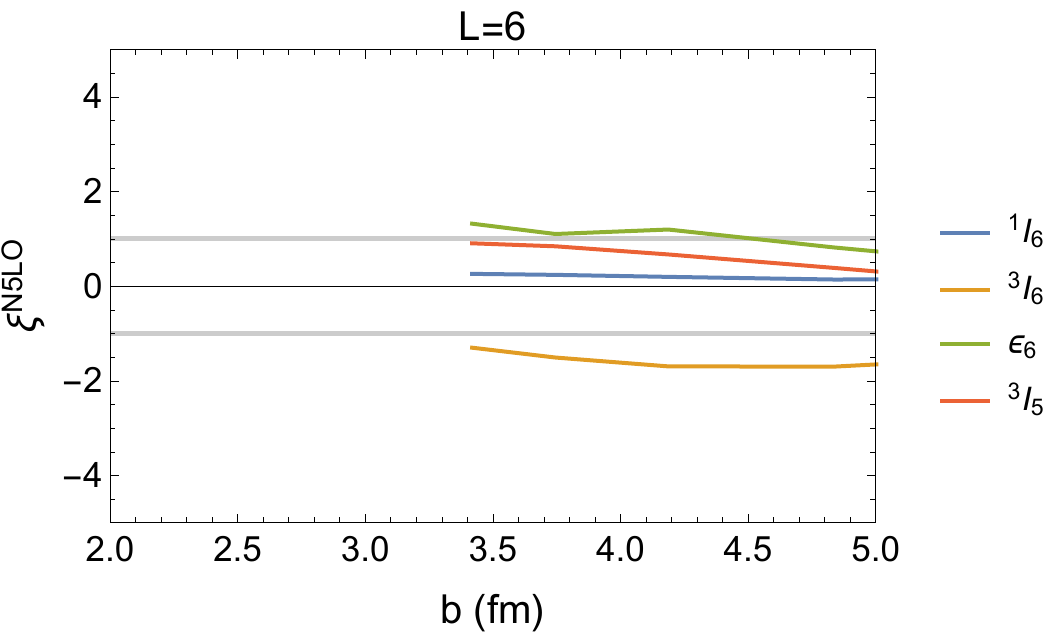} 
\caption{Same as Fig.~\ref{fig:xi-sys-SAID} for the N5LO chiral
  perturbative calculation of Entem {\em et al.,} \cite{Entem:2014msa}.}
\label{fig:xi-sys-N5LO}       
\end{figure*}

\section{Peripheral Tests and Outliers}
\label{sec:peripheral-test}

The previous statistical and systematic error bands set very tight
constraints on any analysis of peripheral waves and we can use them as
a test on the quality of a given interaction and/or scattering
amplitude. We will discuss the peripheral plot in two different cases
for their particular significance and popularity in quite different
contexts: the SAID partial wave analysis \cite{Workman:2016ysf} and
the recent N5LO calculation of Entem {\em et al.,}
\cite{Entem:2014msa}. We will carry out the discussion by conveniently
zooming Figs.~\ref{fig:eik-stat} and \ref{fig:eik-sys} by defining a
quantitative estimate for a given partial wave.  We will consider
three different comparisons.  Firstly, we compare both analyses to the
statistical one carried out in the original paper determining the
Granada-2013 database (DS-OPE~\cite{Perez:2013jpa}), which we denote
$\delta_{\rm Gr}$ and $\Delta_{\rm Gr}$ for the phase-shifts and the
scaled phase-shifts, respectively. Finally, we also compare the SAID
and N5LO analyses to the average of the $N=6$ Granada potentials, and
with the average of the $N=13$ high-quality potentials mentioned in
section \ref{sec:peripheral-hq} and plotted in Fig. \ref{fig:eik-sys}.

Thus we define in the first case the
normalized statistical discrepancy 
\begin{eqnarray}
\xi^{i}|_{\rm stat}  = \frac{\Delta^{i} - \Delta_{\rm  Gr}}{\Delta (\Delta_{\rm Gr})} 
= \frac{\delta^{i} - \delta_{\rm Gr}}{ \Delta \delta_{\rm Gr}}  \, , 
\end{eqnarray}
corresponding to Fig. \ref{fig:eik-stat}. 
In the second and third comparisons, we use Eq.~(\ref{eq:mean}) and
Eq.~(\ref{eq:std}) with $N=6$ and $N=13$, respectively, and we define the
normalized systematic discrepancy
\begin{eqnarray}
\xi^{i}|_{\rm sys}  = \frac{\Delta^{i} - {\rm Mean}(\Delta)}{{\rm Std}(\Delta)} 
= \frac{\delta^{i} - {\rm Mean}(\delta)}{{\rm Std}(\delta)}  \, , 
\end{eqnarray}
This quantity measures the deviation of the phase-shifts ($\Delta_i$
or $\delta_i$) corresponding to the model $i (=$ N5LO, SAID) with
respect to the averaged phase-shifts of the other analyses (either the
6 Granada potentials or the 13 high-quality potentials), divided by
their standard deviation.  The standardized discrepancies $\xi^i(b)$
will be studied as a function of the impact parameter for the most
peripheral partial waves.

It is well known that for any statistical distribution a confidence
level $p$ can be defined. In the case of a $\xi$ variable following a
normal distribution, the confidence level is the probability that
$\xi$ is larger than a fixed value $\xi_0$
\begin{eqnarray}
p(|\xi|>|\xi_0|) = 1- \int_{-\xi_0}^{\xi_0} dx \frac{e^{-x^2/2}}{\sqrt{2\pi}} \, . 
\end{eqnarray}
In the cases $\xi_0=1,2,3$ one has $p=0.32,0.05,0.01$, corresponding
to $1\sigma$, $2\sigma$ and $3 \sigma$ respectively.  Thus the
probability to obtain a result larger than one-sigma ($\xi > 1$) is
32\%, while the probability of it being larger than three-sigmas ($\xi
> 3$) is only 1\%, so it is statistically very unlikely. Here we
understand this $p$-value as the probability of the corresponding
result {\em not being an outlier}. The situation is pictorially
represented in Fig.~\ref{fig:p-value}.

\subsection{SAID partial wave analysis}

The NN PWA at energies below and well above the pion production
threshold has also a long history and a good example of subsequent
upgrades is represented by the series of works conducted by Arndt and
collaborators~\cite{Arndt:1982ep,Arndt:1986jb,Arndt:1992kz,Arndt:1997if,Arndt:2000xc,Arndt:2007qn}
(see also the GWU database~\cite{SAID}). The most recent GWU fit
\cite{Workman:2016ysf}, called SM16, is based on a
parameterization~\cite{Arndt:1986jb} with a total number of 147
parameters, fitting all partial wave amplitudes (phases and
inelasticities) up to $J=7$, below 3 GeV and 1.3 GeV of Lab energy for
pp and np scattering, respectively. It deals with a large body of
$25362$-pp data (with $\chi^2= 48780.9$) and $13033$-np data (with
$\chi^2= 26261.0$), which is sufficient for our considerations
here. This database is probably the largest one and no data selection
was undertaken.

The GWU PWA was carried out up to Lab energy of 3 GeV for the pp case
and hence much smaller impact parameters are reached than with the
analyses stopping at 350 MeV. On the other hand, this analysis does
not incorporate charge dependence in the OPE, nor the small but
crucial electromagnetic effects.

In Fig.~\ref{fig:xi-sys-SAID} we plot the normalized discrepancy
$\xi^{\rm SAID}$ to the original DS-OPE Granada fit (upper panel), to
the average of the six Granada-2013 fits (middle panel) and to the
average of the 13 high-quality potentials (bottom panel).  We only
show the relevant region $b > 2$ fm. Most of the deviations are larger
than one sigma, and in some cases they are even larger than 2 or 3
sigmas if compared with the Granada-2013 band.  According to these
results, the GWU peripheral phase-shifts are outliers, incompatible
with the most accurate PWA of the Granada-2013 database. If compared
with the full set (bottom panel of Fig. \ref{fig:xi-sys-SAID}), some
of the more peripheral waves become more compatible.

In an earlier work~\cite{Arndt:2000xc}, the SAID analysis was also
carried out in the restricted LAB energy range $0-400$ MeV (pp+np) and
compared to the full energy range results going up to 3 GeV (pp) and
1.3 GeV (np). The $0-400$ MeV fits, called SP40, used 30+27=57
variable parameters corresponding to $\chi^2/{\rm datum}=4398/3454$
and $\chi^2/{\rm datum}=5415/3831$. These solutions turned out not to
be very different from the higher energy fits, called SP00, where the
outcoming quality of the fit in the lower $0-400$ MeV range turned
out not to be very different, namely $4593/3454$ (pp) and $5371/3831$
(np). It was thus concluded that high energy SP00 fits did not degrade
the low energy SP40 solutions. We show for comparison the SAID-SP40
solution corresponding to $0-400$ MeV fits in
Fig.~\ref{fig:xi-sys-SAID400}, and the trends are overall similar to
those found in Fig.~\ref{fig:xi-sys-SAID} for the most recent SM16
solution~\cite{Workman:2016ysf}.

\begin{figure*}
\centering
\epsfig{file=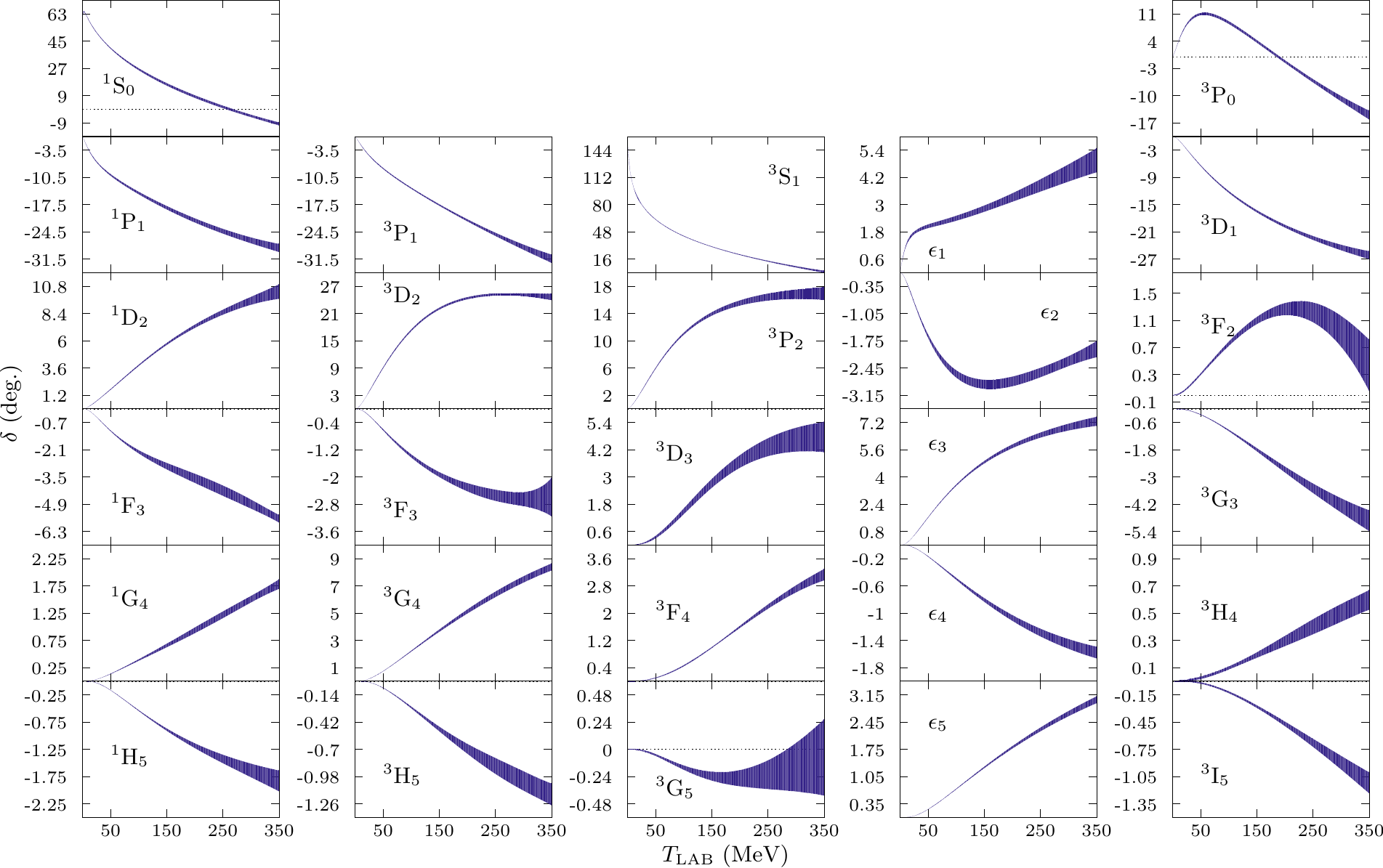,height=11cm}
\caption{Same as Fig.~\ref{fig:ps-sys} but with the band given by $1
  \sigma$ confidence level. See main text.}
\label{fig:ps-sys-1sig}       
\end{figure*}

\subsection{N5LO Chiral Nuclear Forces}

Within a modern perspective and following the proposal by
Weinberg~\cite{Weinberg:1990rz} (see e.g.~\cite{Machleidt:2011zz} for
a comprehensive review) of implementing a power counting based on a
perturbatively rooted effective field theory, chiral symmetry can be
implemented by long distance components described by multiple pion
exchange and short distance components, termed counter-terms, which
are fitted to scattering data. For a given order in the chiral
expansion there is a sufficiently high partial wave where results are
finite and counter-term free\footnote{That means that the short range
  piece of the potential $V_\S (r)$ can be set to zero in that partial
  wave within uncertainties, see e.g. the discussion in
  Ref.~\cite{Perez:2014bua} }.  Within that scheme, peripheral
nucleon-nucleon phase-shifts and chiral symmetry have been
confronted~\cite{Kaiser:1997mw}, and the role of delta excitation,
correlated two pion and vector meson
exchanges~\cite{Kaiser:1998wa,Entem:2002sf} have been analyzed (see
also \cite{Epelbaum:2003gr} and \cite{Krebs:2007rh}). The most recent
bench-marking study on peripheral nucleon-nucleon scattering goes up
to fifth order of chiral perturbation theory~\cite{Entem:2014msa}, and
it confronts to the GWU results. One of the appealing features of the
power counting is that up to a given order in the chiral expansion
peripheral waves are counter-term free. In the case of N5LO this
starts with $F$-waves, i.e. $L \ge 3$. No error bars are included in
the N5LO calculation, although they might be inferred from the
truncation errors in the expansion.

The N5LO results are compared in Fig.~\ref{fig:xi-sys-N5LO} with the
Granada phase-shifts as well as with the 13 potentials considered in
this work sharing the {\it same} CD-OPE potential tail.  As we see
while some N5LO partial waves deviations are within one sigma, some
discrepancies larger than 2--3 sigmas can be also seen, especially for
the most peripheral ones.  However, the N5LO peripheral phase-shifts
become more compatible when the full set of phase-shifts is
considered, as it can be observed from the bottom panels of
Fig. \ref{fig:xi-sys-N5LO}.  Thus, the validation of N5LO requires
admitting that the systematic errors in the phase-shifts are provided
by the spread in the 13 PWA set.

A global view of this spreading of phase-shifts is provided in
Fig.~\ref{fig:ps-sys-1sig} in the conventional representation where we
provide the $1\sigma$ confidence bands computed from
Eqs.~(\ref{eq:mean}) and (\ref{eq:std}). Remarkably, the authors of
the N5LO calculation carried out the comparison with the SAID result
only, which, in view of our plots, is not the best choice.

\section{Conclusions} 
\label{sec:concl}

In the present paper we have conducted a comprehensive study of the
peripheral properties of NN interactions below 350 MeV, focusing in
the design of quantitative tests assessing the long-distance quality
of a given analysis. This approach has the advantage that we may
visualize the effective probing range of the different partial waves
in a fewer number of plots. They have been systematically classified
in the so called peripheral plot.  This has allowed us to find a high
degree of universality in different partial waves sharing the same
CD-OPE potential tail.  This required to compute the scaled
phase-shift $(L+1/2) \delta_L(p)$, which should become an universal
function of the impact parameter $b= (L+1/2)/p$ for large $L$
peripheral waves.  It also implies exploring large values of the
impact parameter, linked to the long range behavior of the NN
interaction.  We also find this peripheral plot to be specially suited
for discussing uncertainties and testing new interactions not based on
a CD-OPE potential and having a high statistical quality.

Using 13 high-quality sets of phase-shifts, starting from the early
Nijmegen-1993 database and covering up to the most recent Granada-2013
one, we have analyzed their scaling properties for large $b$. All
these 13 analyses own their high-quality character to the fact that
they share the same CD-OPE potential in coordinate space and include
all needed electromagnetic effects. If we take the spread of all 13
analyses as a measure of the systematic error, the peripheral scaling
works fairly well.

Finally we have performed three peripheral tests for two recent sets
of phase-shifts, which do not enjoy the high-quality character, based
on the normalized discrepancies with three sets involving high-quality
analyses. The first test corresponds to the original DS-OPE potential
used to select the Granada-2013 database, giving the most accurate fit
to the largest NN-scattering data-set up to date. The second test
comprises the mean and standard deviation with respect to the 6
Granada-2013 potentials. Finally, the third test includes all 13
high-quality fits carried out since the Nijmegen analysis in 1993.

On one hand, we have tested the peripheral structure of the SAID
analysis comprising the largest number of NN scattering data that has
been analyzed to date, and going up to 3 GeV and 1.3 GeV for pp and np
data, respectively. This analysis does not contain the high-quality
long-distance features and, as expected, it fails most of the
peripheral tests. Therefore, the SAID phases are not well-suited for
accurate comparisons below pion production threshold.
 
On the other hand, we consider the recent peripheral analysis within
chiral perturbation theory to fifth order (N5LO) in the
expansion. Here, the peripheral phases have been computed without any
fit to NN scattering, and the couplings entering the calculation are
taken from $\pi N$ scattering. Thus, they can be regarded as pure
predictions of the NN interaction for long distances, hence their
theoretical interest. While we share the view that the rough and
visual agreement with the PWA phase-shifts is already encouraging,
some deviations with respect to the high-quality potentials are large
enough to question the convergence of the expansion. A comparison with
statistical and systematic errors obtained by the latest fits using
the Granada-2013 database provide a complete falsification of these
interactions. The validation of this approach would require assuming
that the dominating errors in the phase-shifts are given by the
average of all 13 high-quality post-Nijmegen 1993 analyses, as
summarized in Fig.~\ref{fig:ps-sys-1sig}.

Our findings have an impact on the predictive power of nuclear
structure calculations and the validation or falsification of nuclear
forces (see e.g. Ref.~\cite{RuizArriola:2016sbf} and references
therein). The basic point is that peripheral waves are not only
determined to much better accuracy than the low-lying partial waves
needed for nuclear structure calculations, but they are essential to
remove systematic errors and to achieve an acceptable statistical
confidence level on the PWA. A failure in the peripheral test beyond
reasonable levels would correspond to an arbitrary enlargement of the
peripheral waves uncertainties with a much larger effect on the low
partial waves.

\section{Acknowledgements}

We thank R. Machleidt for sending us the G, H and I peripheral waves
phase-shifts for the perturbative N5LO chiral interaction and the CD
Bonn ones.

This work is supported by Spanish Ministerio de Economia y
Competitividad and European FEDER funds (Grants No. FIS2014-59386-P
and FIS2017-85053-C2-1-P), the Agencia de Innovacion y Desarrollo de
Andalucia (Grant No. FQM225), the U.S. Department of Energy by
Lawrence Livermore National Laboratory under Contract
No. DE-AC52-07NA27344, U.S. Department of Energy, Office of Science,
Office of Nuclear Physics under Award No. DE-SC0008511 (NUCLEI SciDAC
Collaboration). I.R.S. acknowledges financial support from MINECO
(Spain) under contract No. IJCI-2014-20038 (Juan de la
Cierva-Incorporacion program).

\appendix 

\section{Convergence of the partial wave expansion}
\label{sec:PW-conv}

\begin{figure}
\centering
\includegraphics[height=5cm]{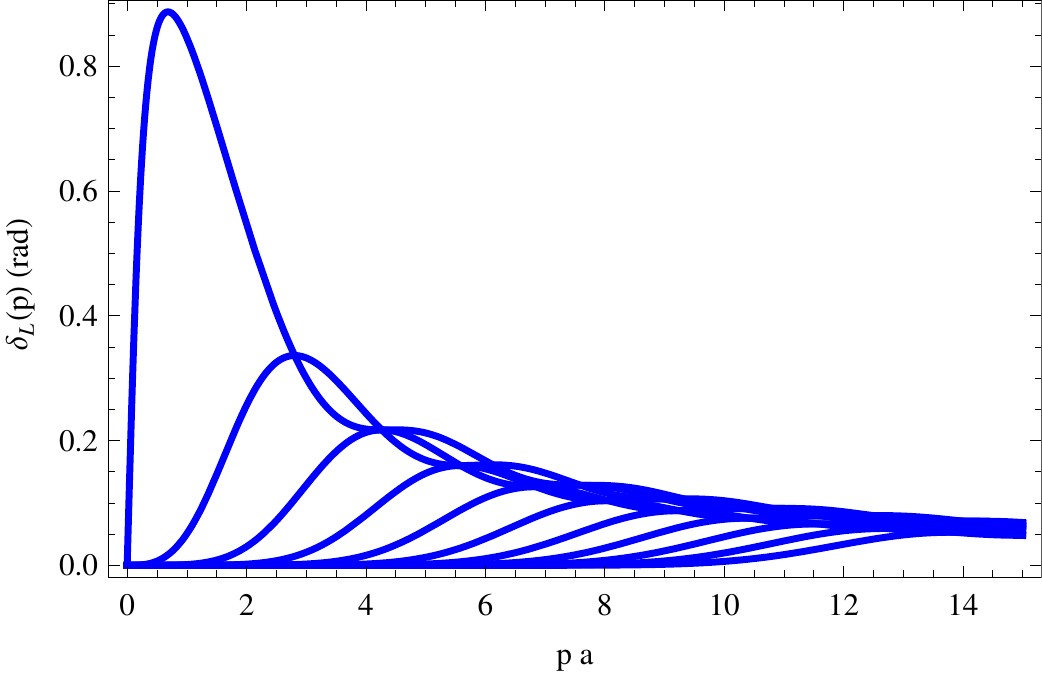}
\includegraphics[height=5cm]{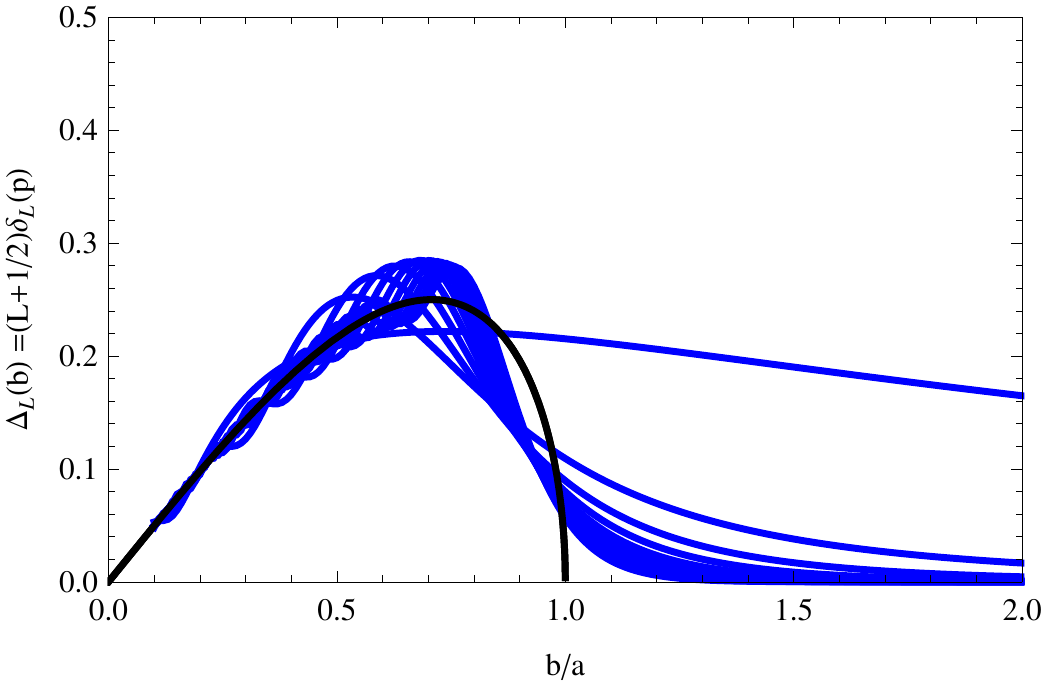} 
\caption{Phase-shifts (upper panel) and the corresponding peripheral
  plot (lower panel) for $l=0, \dots ,10$ (thin solid lines) for an
  attractive spherical well with strength $U_0=2$ and range $a=1$ (in 
  arbitrary units). We also show the eikonal 
  approximation (thick solid black line).}
\label{fig:ps-sw}       
\end{figure}

We analyze the convergence of the partial wave expansion for a
spherical well potential of range $a$ and strength $U_0$.  The
phase-shifts are well-known and they are given by log-matching the
inner and outer reduced wave functions at the point $r=a$.
\begin{eqnarray}
u_{\rm in} (r) &=& \hat j_L \left( \sqrt{p^2 + U_0}r \right) \, \qquad \qquad r < a \, ,  \\
u_{\rm out} (r) &=& \hat j_L ( pr) - \tan \delta_L(p) \hat y_L (p r) \quad \,  r > a \, , 
\end{eqnarray}
 Here $ {\hat j}_L(x) \equiv x j_L(x)$ are the reduced spherical
 Bessel functions, and the same applies for the second kind ones
 $\hat{y}_L(x)$.  A sample result is presented in Fig.~\ref{fig:ps-sw}
 for the phase-shifts and the corresponding peripheral plot compared
 with the eikonal approximation. We see that while $\delta_L(p)
 \approx 0 $ for $L \gtrsim pa $, the peripheral plot shows some
 diffractive effect since $\Delta_L(b)=(L+\frac12)\delta_L(b)$ does
 not vanish at $b \gtrsim a$ but at a somewhat larger value. For
 comparison we also depict the eikonal approximation.

\begin{figure*}
\centering
\includegraphics[height=5cm]{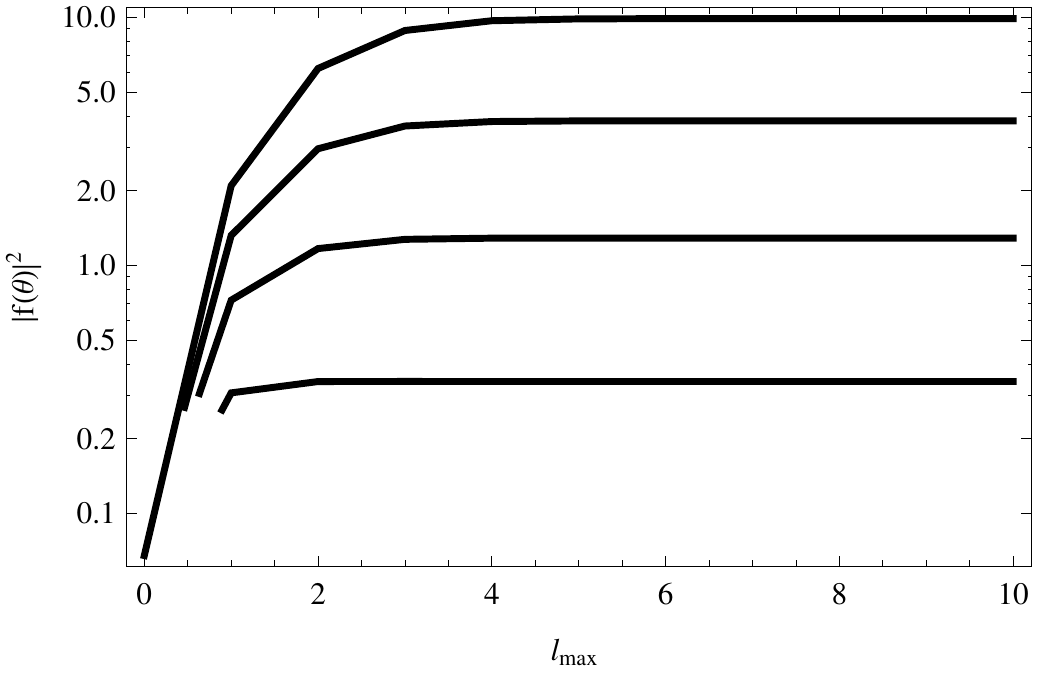}
\includegraphics[height=5cm]{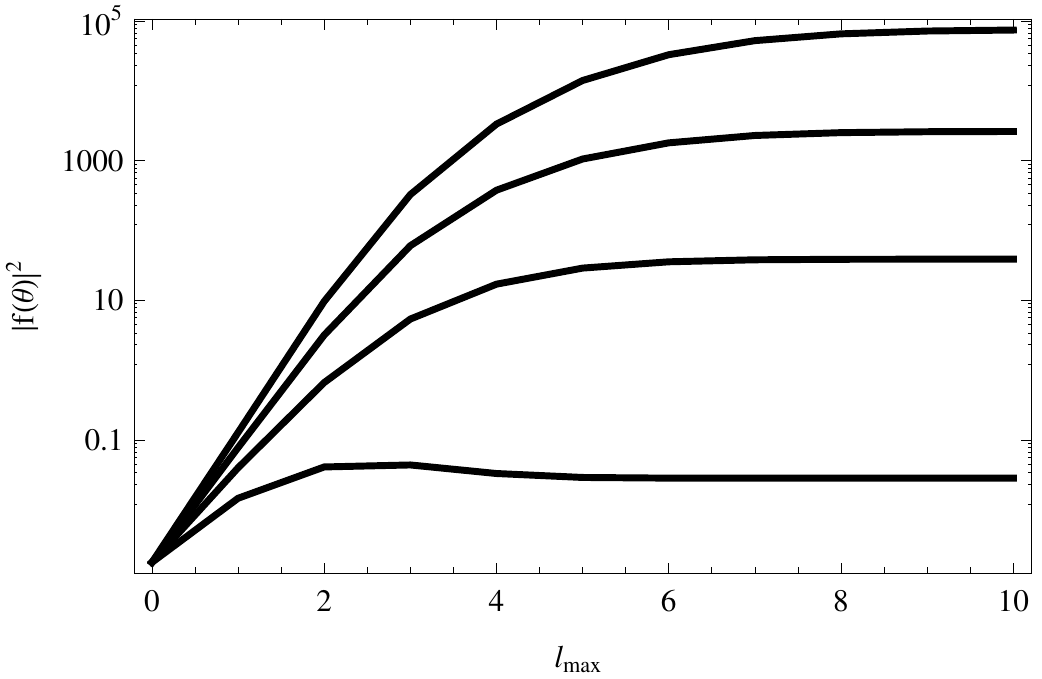}
\includegraphics[height=5cm]{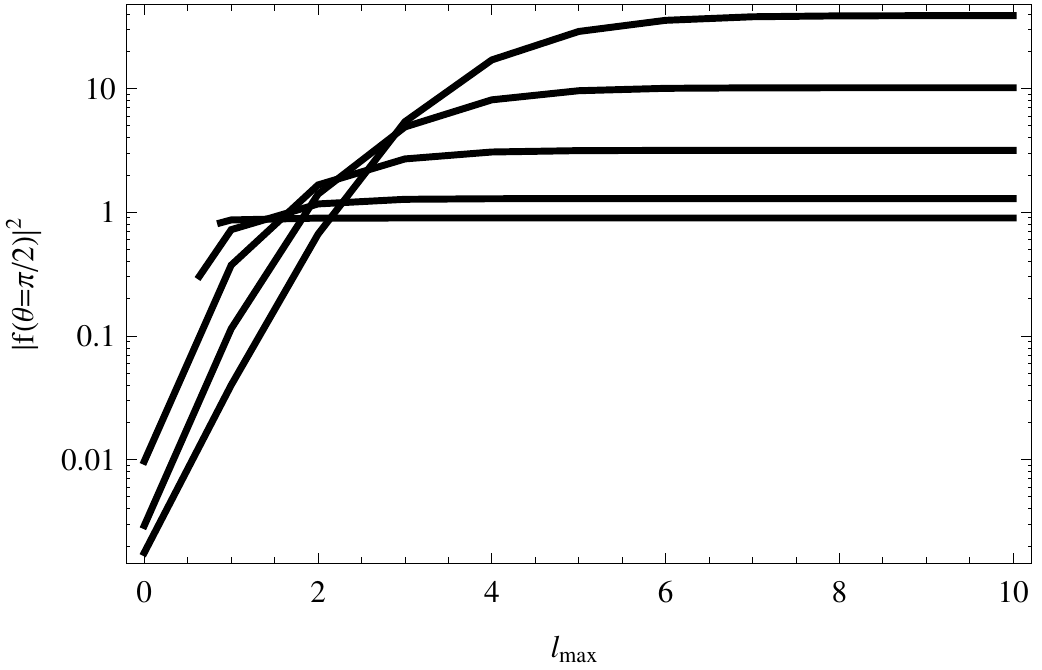}
\includegraphics[height=5cm]{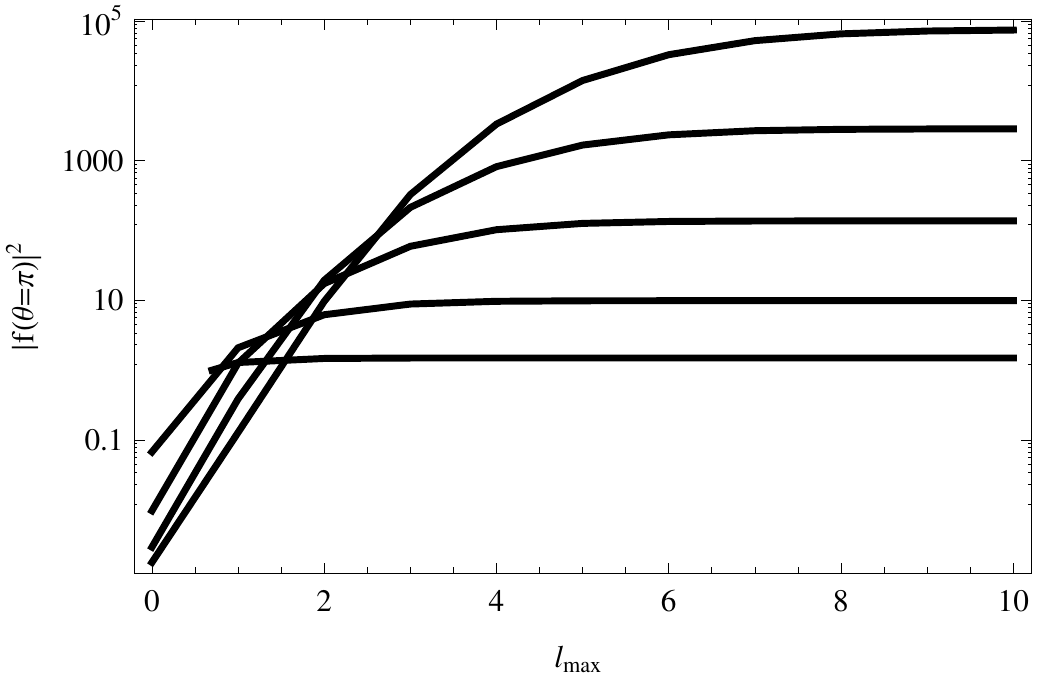} 
\caption{Convergence of the partial wave expansion for an 
attractive  spherical well with strength $U_0=2$ and range $a=1$
(in arbitrary units) as a function of the maximal 
angular momentum $L_{\rm max}$. 
Upper panel: $p=2$ (left) and $p=5$ (right) from top to bottom $\theta=\pi,3\pi/4,\pi/2,\pi/4$. 
Lower panel: $\theta=\pi/2$ (left) and $\theta=\pi$ (right) 
from top to bottom $p=5,4,3,2,1$. }
\label{fig:pwc}       
\end{figure*}

In Fig.~\ref{fig:pwc} we also display the convergence of the partial
wave expansion for several angles and momenta. One can observe that
the maximal value of the angular momentum ($L_{\rm max}$) which is
necessary to be reached in order to saturate the partial wave
expansion, depends on the scattering angle. Generally, it ranges from
$ L_{\rm max} = p a $ to $ L_{\rm max} = 2 p a $ for backward angles.

\section{Perturbative OPE potential}
\label{sec:pert-ope}

The early perturbative treatment of Wu and
Ashkin~\cite{ashkin1948neutron} (see also \cite{wu1962quantum}) is
rather involved and we proceed here differently. Actually, a direct
derivation of perturbation theory can be easily achieved by starting
with the system of coupled channel radial equations (we omit the $S,J$
dependence)
\begin{eqnarray}
-u''_{p,l}(r) &+& \frac{l(l+1)}{r^2} u_{p,l} (r)+ \sum_{l'} U_{l, l'} (r)u_{p,l'} (r)= p^2 u_{p,l} (r)  \, ,\nonumber \\
\label{eq:coupled_diff}
\end{eqnarray}
where $U_{l',l} = 2\mu V_{l'l}$. We can effectively transform this
into an equivalent system of integral equations
\begin{eqnarray}
u_{p,l} (r) &=& {\hat j}_{l} (pr) + \int_0^ \infty dr' \,G_{l} (r,r')\sum_{l'} 
U_{l,l'} (r')u_{p,l'}(r') \, , \nonumber \\
\label{eq:coupled_int}  
\end{eqnarray}
where $G_{l} (r,r')$ is the Green's function of
Eq.~(\ref{eq:coupled_diff}), given by
\begin{eqnarray}
p G_l(r,r') = {\hat j}_{l} (pr_<) {\hat y}_{l} (pr_>) \, , 
\label{eq:green}  
\end{eqnarray}
with $r_<= {\rm min} \{ r,r'\}$ and $r_>= {\rm max} \{ r,r'\}$.  

In the case of OPE the potential reads 
\begin{eqnarray}
V_{\rm OPE}(r) = \tau_1 \cdot \tau_2 
\left[ \sigma_1 \cdot \sigma_2 W_S (r) +  S_{12} W_T (r) \right] \, , 
\end{eqnarray}
and we have that the $V_{L,L'}^{JS}(r)$ matrix elements are
non-vanishing for $J=L$ and $(S,P)=(0,(-1)^L) $, $J=L$ and
$(S,P)=(1,(-1)^J) $ and $L=J\pm 1$ for $(S,P)= (1,(-1)^{L+1}) $ and
fulfilling $(-1)^{L+S+T}=-1$. These potential matrix elements 
are given by
\begin{eqnarray}
V_{J,J}^{J0}(r) &=& \tau \, (-3) W_S (r) \, \\
V_{J,J}^{J1}(r) &=& \tau \, \left[ W_S (r) + 2 W_T(r) \right] \, , \\
V_{J+1,J+1}^{J1}(r) &=& \tau \,  \left[ W_S (r) - \frac{2(J+2)}{2J+1} W_T(r) \right] \, , \\
V_{J-1,J-1}^{J1}(r) &=& \tau \, \left[ W_S (r) - \frac{2(J-1)}{2J+1} W_T(r) \right]\, ,  \\
V_{J+1,J-1}^{J1}(r) &=& \tau \,   \frac{6 \sqrt{J(J+1)}}{2J+1}W_T(r) \, , 
\end{eqnarray}
where $\tau= \tau_1 \cdot \tau_2= -3,1$ for $T=0,1$, respectively.  

The reaction matrix is given by 
\begin{eqnarray}
R_{l',l}^{JS}(p',p) &=& \frac{1}{p^\prime\, p} \int_0^\infty dr \, 
\hat{j}_{l'}(p'r) 2 \mu V_{l',l}^{JS} (r) u_{p,l}^{JS}(r) \, . \nonumber \\
\end{eqnarray}
In the nuclear bar representation the S-matrix reads, 
\begin{eqnarray}
S^{J1} &=&\left(
\begin{array}{cc}
  e^{i \bar \delta_{J-1,J-1}^{J,1}} & 0 \\
  0 & e^{i \bar \delta_{J+1,J+1}^{J,1}}
\end{array}
\right) \left(
\begin{array}{cc}
 \cos 2 \bar \epsilon_J & i \sin 2 \bar \epsilon_J \\
 i \sin 2 \bar \epsilon_J & \cos 2 \bar \epsilon_J
\end{array}
\right) \nonumber \\
&\times& 
\left( \begin{array}{cc}
  e^{i \bar \delta_{J-1,J-1}^{J,1}} & 0 \\
  0 & e^{i \bar \delta_{J+1,J+1}^{J,1}}
\end{array} 
\right) \, . 
\end{eqnarray}
From here we define the T-matrix
\begin{equation}
S^{JS}= 1 - 2 i p T^{JS} \, ,    
\end{equation}
and the on-shell reaction matrix $R$ (for $p^\prime=p$)
\begin{equation}
R_J^{-1}=  T_J^{-1}- i p    \, . 
\end{equation}
In the limit of small phases, $\bar \epsilon_J, \bar \delta_J \to 0 $, we get  
\begin{equation}
R_J =  - \frac1{p} \left(
\begin{array}{cc}
  \bar \delta_{J-1,J-1}^{JS}  & \bar \epsilon_J \\
  \bar \epsilon_J &   \bar \delta_{J+1,J+1}^{JS} 
\end{array} \right) + \cdots 
\end{equation}
Using this form and the perturbative series for the wave function we
get (again for $p^\prime=p$)
\begin{eqnarray}
R_{l',l}^{JS}(p,p) &=& \frac{1}{p^2}   \int_0^\infty dr \, 
\hat{j}_{l'}(pr)\, 2 \mu V_{l',l}^{JS} (r)\, \hat{j}_{l}(p r) \nonumber \\
&+& {\cal O} (V^2) \, , 
\end{eqnarray}
whence Eqs. (\ref{eq:deltabar_l_j}), (\ref{deltabar_pm}) and 
(\ref{eq:ps-OPE}) follow. The corresponding integrals were
determined in
Refs.~\cite{signell1959nucleon,Cziffra:1959zza,breit1960note} and for
completeness we quote here the corresponding expressions
for the scaled phase-shifts\footnote{There is a typo in
  Ref.~\cite{breit1960note} which, however, has no consequences in the
  final formula}
\begin{eqnarray}
\Delta^{S=0}_J &=& \tau \frac{M p f^2}{2 m^2} 
(2 J+1) (z-1) Q_J(z) \, , \\
\Delta^{S=1}_{L=J} &=& \tau \frac{M p f^2}{2 m^2}\left[ -(J+1) Q_{J-1}(z)+(2
   J+1) Q_J(z)\right. 
   \nonumber\\
   &&\left.-J
   Q_{J+1}(z) \right] \, ,   \\
\Delta^{-}_{SLJ}  &=& \tau \frac{M p f^2}{2 m^2}\left[ Q_{J-1}(z)-Q_J(z) \right] 
\frac{(2J-1)}{(2J+1)} \, , 
 \\
 \Delta^{+}_{SLJ} &=& \tau \frac{M p f^2}{2 m^2}\left[ Q_J(z)-Q_{J+1}(z) \right] 
 \frac{(2J+3)}{(2J+1)} \, , 
 \\
  \Delta^{\epsilon}_{SLJ} &=& - \tau \frac{M p f^2}{2 m^2}
   \sqrt{J(J+1)} \left[Q_{J-1}(z)-2 Q_J(z)\right. 
   \nonumber\\
 &&\left.   +Q_{J+1}(z)\right] \, , 
\end{eqnarray}
where $Q_J (z) $ are the Legendre functions and $z= 1 + m^2/2p^2$.
These functions have branch points at $z = \pm 1$ so that the branch
cut must be specified. Here, we just take the unambiguous real part.

\section{Scaled phase-shifts in the WKB approximation}
\label{sec:WKB}

We proceed by using the WKB representation of the reduced spherical
Bessel functions, as an approximation to the free wave solutions in
the classically allowed region, i.e, for $r > r_0$, with $r_0$ a
turning point of the classical trajectory. This WKB representation can
be written as
\begin{eqnarray}
\hat j_{L,{\rm WKB}} (r) = \sqrt{\frac{p}{p_L (r)}}\sin \left[ \int_{r_0}^r 
p_L (r^\prime)\, dr^\prime +  \frac{\pi}{4}\right] \, , \label{eq:WKB_bessel}
\end{eqnarray}
where the free local momentum is only modified by the centrifugal 
barrier, and tends to the asymptotic momentum $p$ for
$r\rightarrow\infty$. Its expression is given by
\begin{eqnarray}
p_L (r) = \sqrt{p^2-\frac{(L+1/2)^2}{r^2}} \equiv 
p \sqrt{1-\frac{b^2}{r^2}} \, , \label{eq:local_mom}
\end{eqnarray}
where the semi-classical expression $pb=L+\frac12$ has been used to
introduce the impact parameter in the last step of
Eq. (\ref{eq:local_mom}). Thus, the classical turning point, $r_0$,
can be naturally identified with the impact parameter $b$, as this is
the point on the classical trajectory that satisfies $p_L(r_0)=0$.  In
Eq. (\ref{eq:local_mom}) we have also included Langer's modification
$L(L+1) \to (L+1/2)^2$, in order to implement the correct short
distance asymptotic behavior in the classically forbidden region $r <
b$.

The integral of the local momentum appearing in
Eq. (\ref{eq:WKB_bessel}) can be identified with the classical action
and its result is
\begin{eqnarray}
S_L(r) &=& \int_{b}^r dr^\prime\, p\, \sqrt{1-\frac{b^2}{r^{\prime\,2}}}=   \nonumber\\
&=&p\sqrt{r^2-b^2} - p\,b\, \arccos\left(\frac{b}{r}\right) \, . 
\end{eqnarray}

Finally, to obtain Eq. (\ref{deltapmWKB}) for the scaled phase-shifts
in the WKB approximation, it is necessary to substitute the WKB
representation of the free wave solutions, given by
eq. (\ref{eq:WKB_bessel}), into Eqs. (\ref{eq:deltabar_l_j}) and
(\ref{deltabar_pm}) for the diagonal phase-shifts
\begin{eqnarray}
\left.\bar{\delta}^{\pm}_{SLJ}(p)\right|_{\rm WKB} &=&
-\frac{M}{p} \int_0^{\infty} dr\; V^{J}_{J\pm1,J\pm1}(r)
\frac{r}{\sqrt{r^2-b^2}}\nonumber\\
 &\times&\sin^2\left(S_L(r) +\frac{\pi}{4} \right). \label{eq:wkb_appendix}
\end{eqnarray}
The last step corresponds to use the standard WKB rule to replace the
square of the oscillating function by its average value $\sin^2 x \to
1/2 $, which finally yields Eq.~(\ref{deltapmWKB}) for the scaled
phase-shifts. Note also that the lower limit in the integral
(\ref{eq:wkb_appendix}) is changed from $0$ to $b$ in order to make
sense. Indeed, this is related to the fact that the WKB approximation
to the free wave solution, Eq. (\ref{eq:WKB_bessel}), is only valid in
the classically allowed region, i.e, for $r > b$.

In the case of the mixing phase-shift, notice that
Eq. (\ref{eq:ps-OPE}) corresponds to two different Bessel
functions. When substituting the WKB representation of the free
solutions, Eq. (\ref{eq:WKB_bessel}), the product of two sine
functions with different classical actions arise.  Then, we can use
the trigonometric identity
\begin{eqnarray}
\sin \left[S_{J-1} +\frac{\pi}{4} \right] 
\sin \left[ S_{J+1} + \frac{\pi}{4} \right] 
&=& \frac12 \cos [S_{J-1} - S_{J+1}] \nonumber\\
&-& \frac12 \cos [S_{J-1} + S_{J+1} +\frac{\pi}{2}] \nonumber\\ 
\end{eqnarray}
The sum of actions in the limit of large $J$ is a large phase and
produces an oscillating cosine function whose average value under the
integral sign is zero.
\begin{equation}
S_{J+1}+ S_{J-1} \approx 2 S_J.
\end{equation}

 The difference of actions, however, yields a finite contribution
\begin{eqnarray}
\cos \left[S_{J-1}-S_{J+1}\right] &\approx&  
\cos\left[2\arccos\left( \frac{b}{r} \right) \right]=
\frac{2b^2-r^2}{r^2}\; ,\nonumber\\
\end{eqnarray}
which makes sense if $r>b$.

In this way, the product of two reduced spherical Bessel functions
with different orders can be approximated under the integral sign by
\begin{eqnarray}
&&\sqrt{\frac{p}{p_{J-1}}} \sin\left[ S_{J-1}+ \frac{\pi}{4} \right] 
\sqrt{\frac{p}{p_{J+1}}}
\sin\left[ S_{J+1}+ \frac{\pi}{4} \right] \to \nonumber\\ 
&\to&
\frac{p}{p_J(r)}\, \frac{2b^2-r^2}{2r^2} =  \frac{2 b^2-r^2}{2 r \sqrt{r^2-b^2}} \, , 
\end{eqnarray}
and then the scaled mixing phase-shift 
$\Delta^{\epsilon}_{SLJ}(b)_{\rm WKB}$, given by Eq. (\ref{deltaeWKB}), 
is obtained.


\end{document}